\documentclass[sigconf]{acmart}

\usepackage{graphicx}
\usepackage{subfig}

\usepackage{paralist}
\usepackage{tabularx}

\usepackage{lipsum}
\usepackage{soul,color}

\usepackage{listings}
\usepackage{xcolor}

\colorlet{punct}{red!60!black}
\definecolor{background}{HTML}{EEEEEE}
\definecolor{delim}{RGB}{20,105,176}
\colorlet{numb}{magenta!60!black}

\lstdefinelanguage{json}{
    basicstyle=\normalfont\ttfamily,
    numbers=left,
    numberstyle=\scriptsize,
    stepnumber=1,
    numbersep=8pt,
    showstringspaces=false,
    breaklines=true,
    frame=lines,
    backgroundcolor=\color{background},
    literate=
     *{0}{{{\color{numb}0}}}{1}
      {1}{{{\color{numb}1}}}{1}
      {2}{{{\color{numb}2}}}{1}
      {3}{{{\color{numb}3}}}{1}
      {4}{{{\color{numb}4}}}{1}
      {5}{{{\color{numb}5}}}{1}
      {6}{{{\color{numb}6}}}{1}
      {7}{{{\color{numb}7}}}{1}
      {8}{{{\color{numb}8}}}{1}
      {9}{{{\color{numb}9}}}{1}
      {:}{{{\color{punct}{:}}}}{1}
      {,}{{{\color{punct}{,}}}}{1}
      {\{}{{{\color{delim}{\{}}}}{1}
      {\}}{{{\color{delim}{\}}}}}{1}
      {[}{{{\color{delim}{[}}}}{1}
      {]}{{{\color{delim}{]}}}}{1},
}
\acmConference{arxiv}{draft}{}

\title{adPerf: Characterizing the Performance of Third-party Ads}

\author{Behnam Pourghassemi, \space Jordan Bonecutter, \space Zhou Li, \space Aparna Chandramowlishwaran}

\affiliation{\institution{University of California, Irvine}  \city{Irvine}
  \state{CA}
  \country{USA}}

\setcopyright{none}
\renewcommand\footnotetextcopyrightpermission[1]{}
\begin{document}

\begin{abstract}
Monetizing websites and web apps through online advertising is widespread in the web ecosystem, creating a billion-dollar market.
This has led to the emergence of tertiary ad providers and ad syndication that facilitate this growing market.
The online advertising ecosystem nowadays forces publishers to integrate ads from these third-party domains.
On the one hand, this raises several privacy and security concerns that are actively studied in recent years.
On the other hand, given the ability of today's browsers to load dynamic web pages with complex animations and Javascript, online advertising has also transformed and can have a significant impact on webpage performance.
The performance cost of online ads is critical since it eventually impacts user satisfaction as well as their Internet bill and device energy consumption.
Unfortunately, there are limited literature studies on understanding the performance impacts of online advertising which we argue is as important as privacy and security.

In this paper, we apply an in-depth and first-of-a-kind performance evaluation of web ads.
Unlike prior efforts that rely primarily on adblockers, we perform a fine-grained analysis on the web browser's page loading process to demystify the performance cost of web ads. 
We aim to characterize the cost by every component of an ad, so the publisher, ad syndicate, and advertiser can improve the ad's performance with detailed guidance.   
For this purpose, we develop an infrastructure, \emph{adPerf}, for the Chrome browser that classifies page loading workloads into ad-related and main-content at the granularity of browser activities (such as Javascript and Layout).
Our evaluations show that online advertising entails more than 15\% of browser page loading workload and approximately 88\% of that is spent on JavaScript.
We also track the sources and delivery chain of web ads and analyze performance considering the origin of the ad contents.
We observe that $2$ of the well-known third-party ad domains contribute to 35\% of the ads performance cost and surprisingly, top news websites implicitly include unknown third-party ads which in some cases build up to more than 37\% of the ads performance cost.
\end{abstract}


\maketitle
\pagestyle{plain}

\section{Introduction}
Online advertising has proliferated in the last decade to the extent where it is now an integral part of the web ecosystem.
Today, publishers display one or multiple advertisements (or ads) through pop-ups, banners, click-throughs, iframes, widgets, etc, to monetize their websites and web apps.
The majority of these ads neither come from the publisher (website) nor a specific domain.
They are delivered through a chain of third-party content providers (such as ad providers, syndication agencies, ad exchange traders, trackers, and analytics service providers) who are part of a complex ad network on the server-side \cite{ikram2019chain}.
The current ad delivery method forces publishers to embed unknown third-party content (such as JavaScript or HTML) on their website which could jeopardize user privacy and security.
There have been several studies in recent years to locate the untrusted sources and/or malicious ad contents \cite{li2012knowing, bashir2016tracing, ikram2019chain, trackingStudy, oneMilion, brokers}.
Accordingly, different blocking and evasion policies have been devised to guard against such malware and aggressive tracking \cite{ bashir2016tracing, iqbal2018adgraph}.

While user privacy and security are of paramount importance, it is not the solitary concern of the worldwide web community.
Online advertising also has a direct impact on website performance (eg., page load time) and in turn user satisfaction.
According to Google, 53\% of mobile site visitors leave a page that takes longer than three seconds to load~\cite{An:2018aa}.

\begin{figure} [htbp]
\def\tabularxcolumn#1{m{#1}}
\begin{tabularx}{.5\textwidth}{cc}

\subfloat[\texttt{ebay.com} in 2002]{\includegraphics[width=0.22\textwidth]{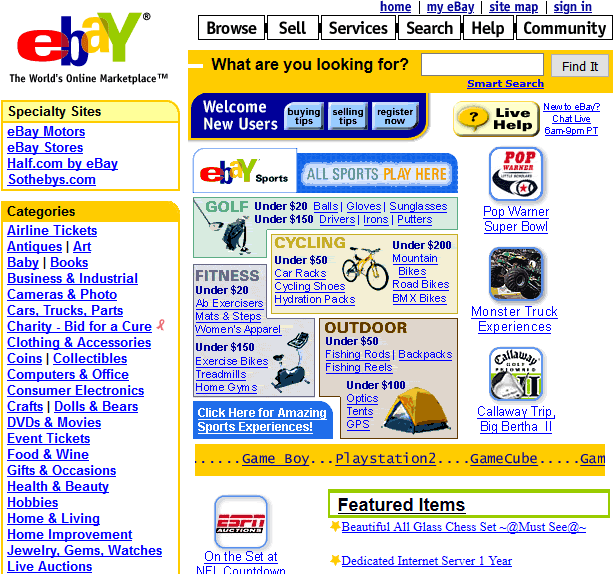}}
&
\subfloat[\texttt{ebay.com} in 2020]{\includegraphics[width=0.23\textwidth]{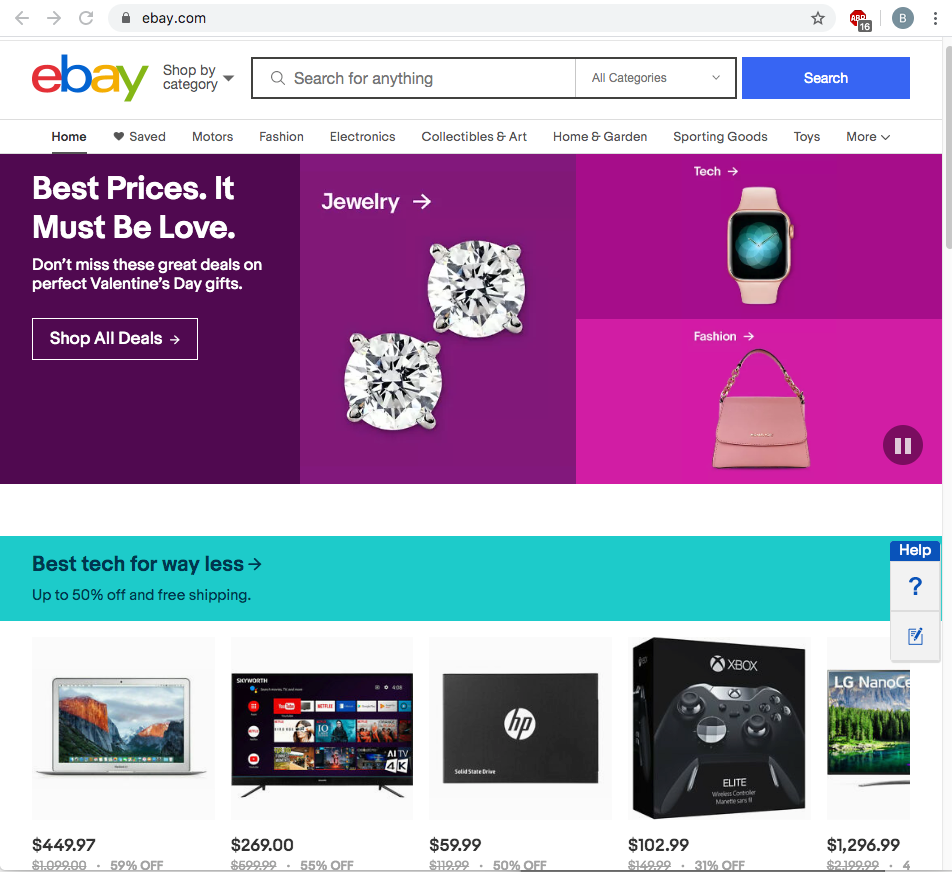}}

\end{tabularx}

\caption{Evolution of ads on the web. (a) Early web ads contain text, image, and hyperlink. (b) Today's complex and dynamic web ads (rotating on top of the website) contain JavaScript, animation, multimedia, and iframe.}\label{fig:ebay}
\end{figure}

Web ads have become more diverse and complex keeping up with the pace of advances in web design.
Figure \ref{fig:ebay} compares advertising on \texttt{ebay.com} in 2002 and 2020.
As we can observe, in the past, ads only included hypertext and images.
However, today's online ads comprise of JavaScript, iframe, animation, multimedia, etc.
Therefore, evaluating and displaying these dynamic ad contents demand increased computation from the browser.
Coupling this observation with recent studies \cite{Pourghassemi:2019aa, wprof, fastlayout} that show that most of the page load time is spent on computation activities in modern browsers raises several intriguing questions.
For instance, (a) \emph{how much do ads increase the browser's page rendering workload?} and (b) \emph{what type of ad contents contribute more to the performance of the website?}
The performance overhead of loading intrusive ads become even more important since it also has an impact on the energy consumption of the device and the user's Internet bill.
Therefore, understanding how much ads contribute to the breakdown of different activities in modern browsers can inform the design of efficient ads and/or optimizations targeting those specific activities.
Unfortunately, only a handful of studies \cite{annoyedUsers, ikram2017first, adblockPerformance} have been devoted to the performance analysis of ads, yet many such important open questions remain to be answered.

Previous studies revolving around the performance analysis of ads lack a comprehensive examination for at least the following reasons.
First, the majority of them concentrate on the network data traffic overhead, neglecting the performance cost of browser computation activities such as rendering activities \cite{adblockPerformance, annoyedUsers}.
Second, prior efforts fundamentally share the same approach for quantifying the performance of ads.
They use ad-blockers to block websites' ad contents and assess the performance overhead via comparison with \emph{vanilla} run (no ad-blocking).
This approach, however, is prone to inaccuracy as it does not take into account the intrinsic overhead of the ad-blocker.
Our measurement over 350 websites shows Adblock Plus \cite{adblockplus}, the most popular and optimized ad-blocker today, adds 32\% overhead (median) to the page loading due to exhaustive filter rule matching even though it ultimately reduces page load time by aggressive content blocking.
Furthermore, ad-blockers are known to lead to site breakage and undesired app functionality, particularly, with the prevalence of anti-ad-blockers \cite{ikram2017first,iqbal2017ad}.
Finally, the aforementioned approach suffers from an inability to conduct comprehensive and fined-grained performance analysis.
This is because ad-blockers block ad-related content as early as the initiation of network requests.
Thus, subsequent ad-related activities such as content parsing and rendering and descendent resource loading remain invisible for inspection.

In this paper, we extensively investigate the performance overhead of all types of ad-related content by crawling over 500 websites.
Unlike previous efforts, we take a novel approach based on in-browser profiling that does not rely on ad-blockers.
Our proposed methodology enables the browser to automatically fetch and evaluate ads' performance at scale.
It correlates the browser's computation and network activities to the associated ad contents and quantifies the added cost of loading ads.
We break down the performance overhead to individual requests and content types through a novel resource mapping technique.
This procedure contrives a more robust and detailed performance analysis. 
Moreover, we demystify and track down ad components on the publisher and characterize the performance overhead considering the origin of ads and how they are delivered to the publisher.
To the best of our knowledge, this is the first time such an experiment has been conducted.

\textbf{Contributions and Findings.} To summarize, this paper makes the following contributions.

\begin{itemize}
\item We employ a different yet more appropriate methodology to characterize the performance overhead of ads.
Our method avoids using ad-blockers, providing higher accuracy and capability for fine-grained measurements while suppressing site breakages and app failures observed in prior studies.
We implement an infrastructure called \emph{adPerf} based on our proposed approach for the Chrome browser since it the most commonly used browser by desktop and mobile users.
The key challenge we encountered is how to align the performance cost with individual components within an ad (e.g., image and JavaScript code), and we address this through a carefully designed resource mapper.

\item Using adPerf, we perform an in-depth and comprehensive evaluation to demystify and locate the performance cost of web ads.
We crawl and analyze over 500 websites from different categories.
Our large-scale examination leads to several first-of-a-kind findings that shed light on the performance cost of ads, giving website builders and web ad providers deeper understanding to mitigate the performance penalty of ads.
For example, our results show that on average 15\% of browser page loading activities are spent on ad-related content for Alexa top 350 news websites.

\item To perform a detailed source-to-target analysis of web ads, we construct the dependency graph for the website's resources and track the delivery chain involved in third-party ads.
The results show that \texttt{googletagservices.com} and \texttt{doubleclick.net}, two reputable ad domains, contribute 35\% of the ad resources resulting in the largest performance cost of online advertising.
Moreover, we characterize the trustworthiness and prevalence of third-party ad domains and distinguish the performance overhead of such domains on the web ecosystem.
Almost half of the highly-visited websites implicitly trust uncommon third-party ad domains and our results show that about 37\% of ads performance cost is related to untrusted ad domains.

\item We will release the source code of adPerf and the detailed measurement results of each website~\footnote{will be available at https://gitlab.com/adPerf/adPerf}.

\end{itemize}

The rest of the paper is structured as follows.
We discuss related work in section \ref{sec:related} followed by essential background in section \ref{sec:background}.
Our methodology to characterize performance and detailed description of the design and implementation of adPerf is presented in section \ref{sec:adperf}.
We discuss the experimental setup in section \ref{sec:setup} and present our results along with a discussion of several findings in section \ref{sec:results}.
The summary of this paper along with key takeaways are presented in section \ref{sec:conc}.
\section{Related work}
\label{sec:related}
Online advertising (essentially display ads on websites) has been rapidly growing in the last decade, generating a multi-billion dollar market \cite{onlineads, onlineads1}.
In the past few years, the focus of academic research has been primarily centered on detecting malicious ad contents (malware and aggressive tracking)~\cite{li2012knowing, bashir2016tracing, ikram2019chain, trackingStudy, oneMilion, brokers} and blocking them~\cite{iqbal2018adgraph, zhu2019shadowblock, tackingthetrackers}.
While user privacy and security are of paramount importance, even ads that are safe and not tracking users can have significant performance impact which in turn causes cascading effects on user satisfaction and Internet costs.

\subsection{Performance analysis of ads}
In a 2015 study, the New York Times analyzed the top 50 news websites landing page containing both advertising and editorial content and found that more than half of all data came from ads~\footnote{https://www.nytimes.com/interactive/2015/10/01/business/cost-of-mobile-ads.html}.
For instance, \texttt{Boston.com}'s mobile website ads took 31 seconds to load on a typical 4G connection while the editorial content only took 8 seconds.
This is equivalent to 32 cents of cell data in ads every time the landing page is loaded.
However, this impact of online advertising on webpage performance (i.e., page load time) is still not well studied.

The few notable efforts similar to the New York Times study, rely on adblockers to measure the performance impact of ads.
Pujol et al~\cite{annoyedUsers} crawled the Alexa top 1k sites and find that 18\% of total traffic in a residential broadband network was due to ads.
However, this does not account for the actual performance of the adblockers themselves.
Ikram et al~\cite{ikram2017first} analyzed 97 ad-blocking apps on Android and reported that 7\% of user complaints relate to crashes and performance-related aspects such as battery-life overhead.
Garimella et al.~\cite{adblockPerformance} analyze the performance of several popular ad-blockers such as Adblock Plus, Ghostery, uBlock, etc.
They conclude that in some cases the time to load pages is not faster when using adblockers since they contribute to additional overhead due to various tracking services of their own.

The key distinction between this paper and prior efforts is that we do not rely on ad-blockers for performance analysis of ads for three main reasons.

\begin{figure} [htbp]
\centering
\includegraphics[width=0.95\linewidth]{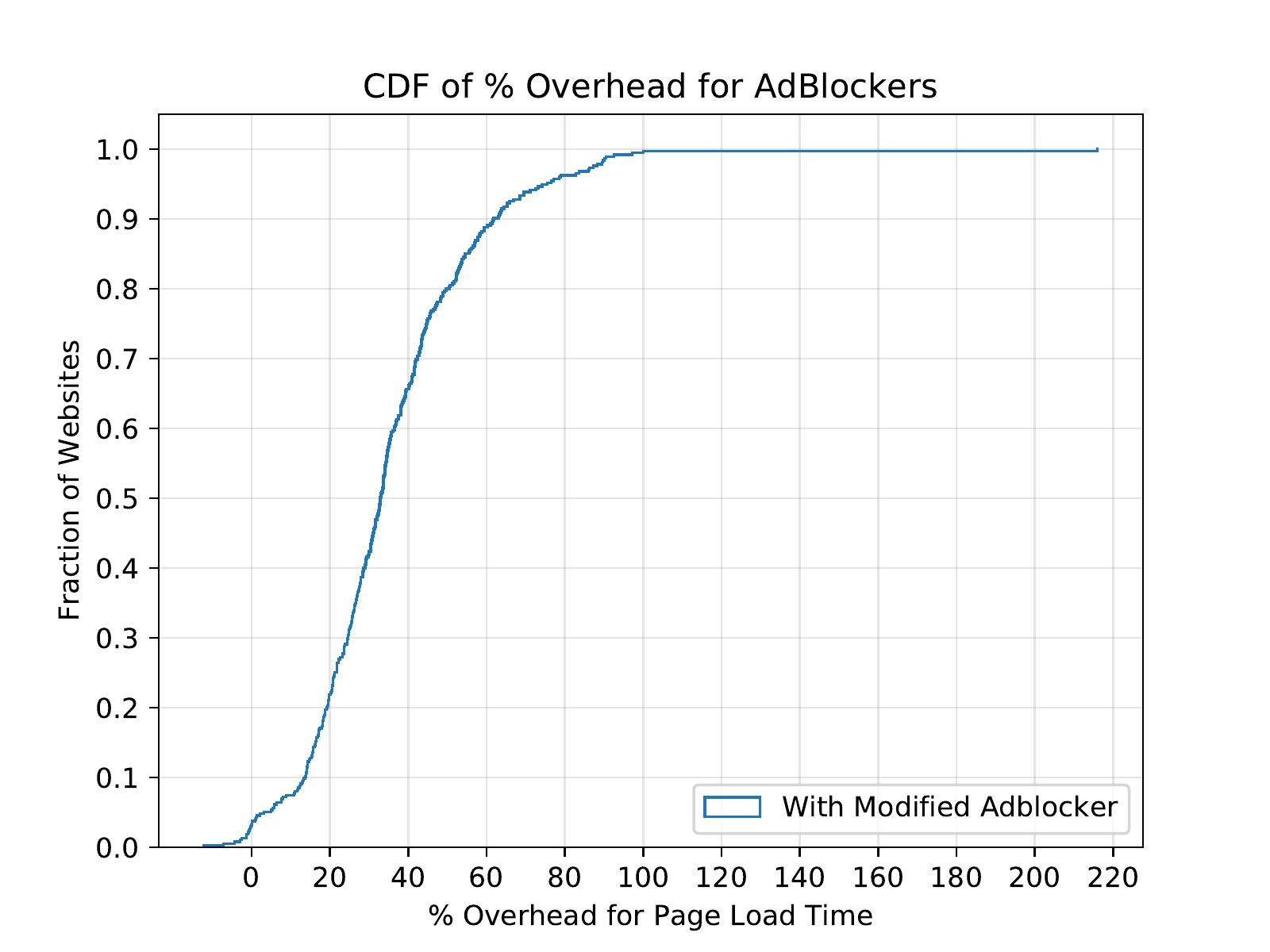}
\caption{CDF distribution of AdblockPlus overhead on the page loading of 350 webpages.}
\label{fig:adblocker-overhead}
\end{figure}

\textbf{\emph{Overhead.}} Similar to the above studies, our results show that ad-blockers themselves can have significant performance overhead due to exhaustive filter-list matching and tracking services.
We analyzed Adblock Plus by creating a modified version that still performs all of the content filtering operations without actually blocking any of the content.
We calculate the overhead imposed by these filtering operations by measuring the difference in page load times from the modified version of Adblock Plus to Vanilla Chromium.
Figure \ref{fig:adblocker-overhead} shows the overhead of Adblock Plus on 350 webpages.
According to the figure, for half of the websites, Adblock Plus adds more than 32\% overhead to the page loading due to excessive and CPU-intensive filter rule matching.
However, it ultimately reduces page load time by aggressive content blocking.

\begin{figure} [htbp]
\centering
\includegraphics[width=0.9\linewidth]{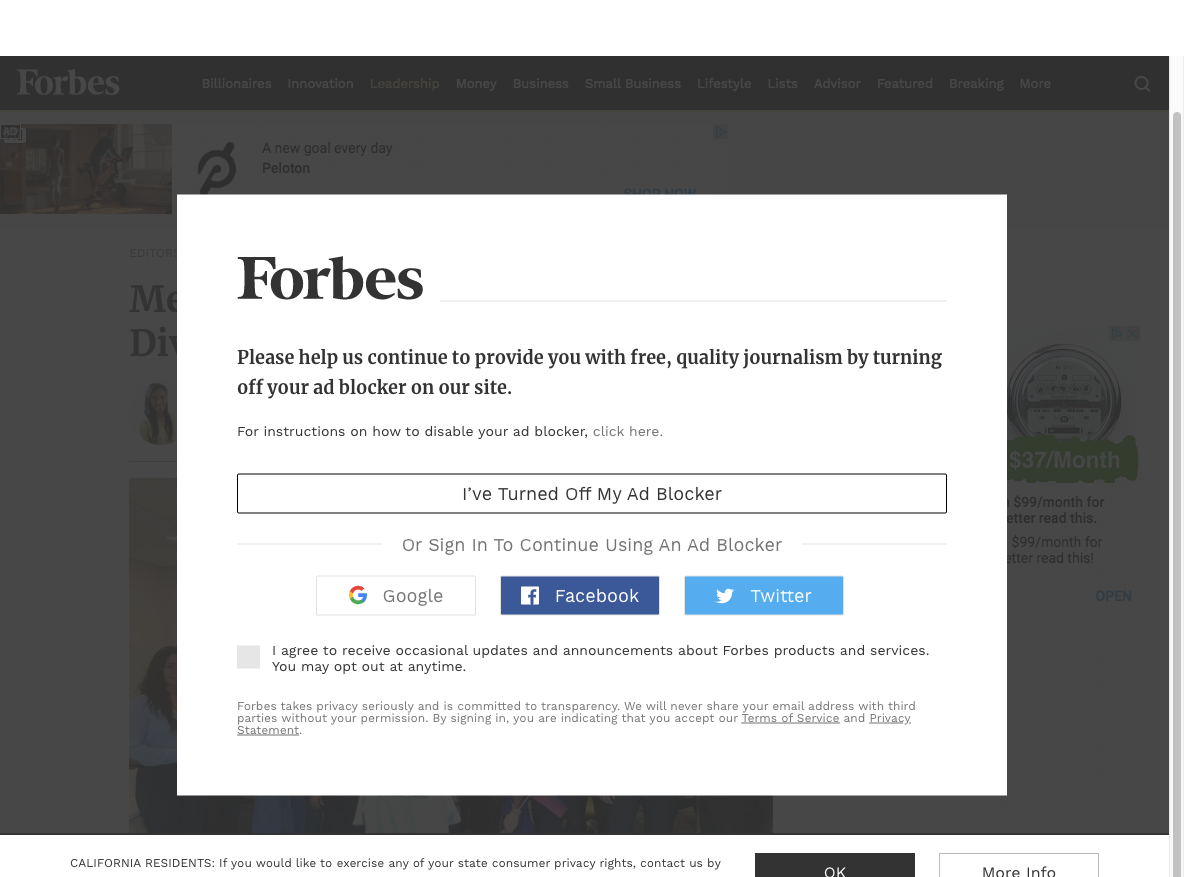}
\caption{Snapshot of \texttt{www.forbes.com}. This website prevents loading contents if visitors attempt to block ads.}
\label{fig:forbes}
\end{figure}

\textbf{\emph{Functionality.}} As ad-blockers become a threat to the "free" web business model, many websites prevent displaying their content to the visitors that use ad-blockers.
In this case, the publisher includes a script such as IAB ad block detection script~\cite{iab} that monitors the visibility of ads to DEAL (Detect, Explain, Ask, Limit) with ad-blockers ~\cite{mughees2017detecting}.
Typically, when the publisher detects a hidden or removed ad, it stops loading the website by displaying a popup that asks the visitor to turn off the ad-blocker.
Figure \ref{fig:forbes} shows a snapshot of the content-blocking of \texttt{www.forbes.com} when ad-blocker is on.

\begin{figure} [htbp]
\def\tabularxcolumn#1{m{#1}}
\begin{tabularx}{.5\textwidth}{cc}

\subfloat[before content blocking]{\includegraphics[width=0.22\textwidth]{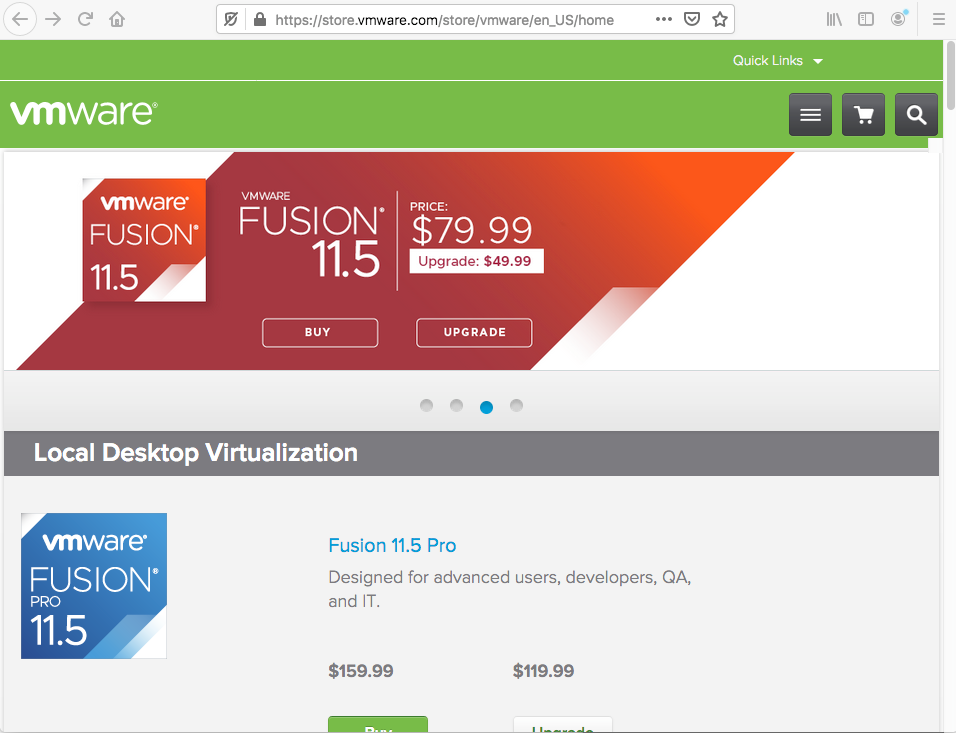}}
&
\subfloat[after content blocking]{\includegraphics[width=0.23\textwidth]{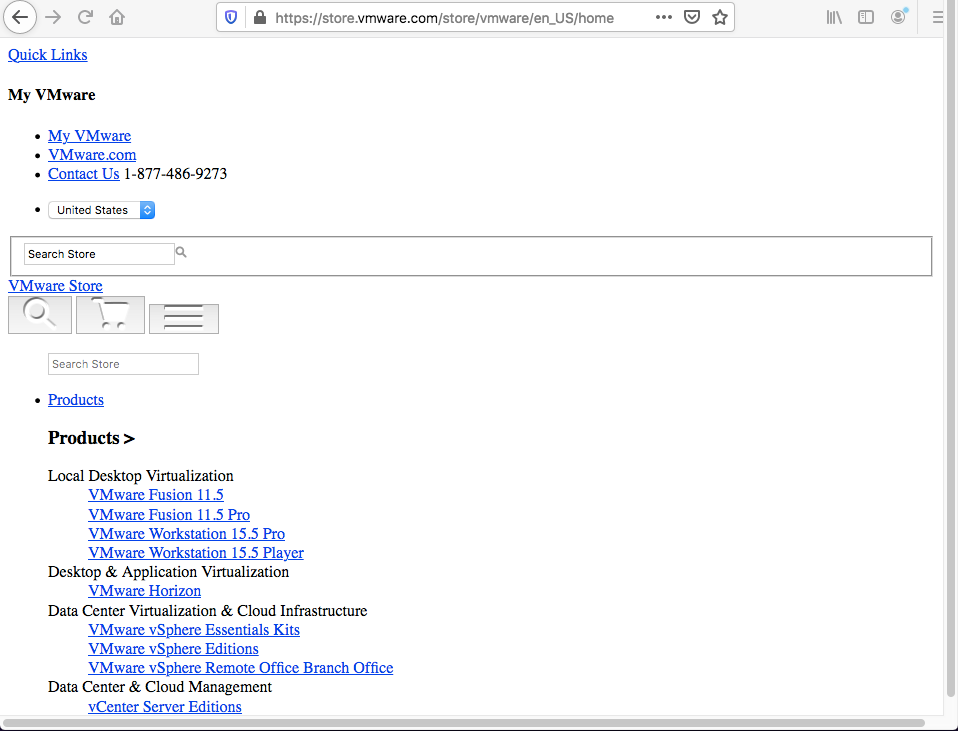}}

\end{tabularx}

\caption{Snapshots of \texttt{www.store.vmware.com}. The layout of the page is broken due to content blocking. }\label{fig:break}
\end{figure}

Content-blocking can lead to site breakage and also other undesired app functionality ~\cite{iqbal2017ad}.
This breakage can range from a dysfunctionality in part of the website (e.g. not displaying login popup) to the breakdown of the entire website layout.
Figure \ref{fig:break} shows a snapshot of \texttt{www.store.vmware.com} when Mozilla's ad and tracking protection is turned on.
Furthermore, a large number of websites employ ad-blocking circumvention to evade from adblocking.
For instance, \texttt{www.thoughtcatalog.com} and \texttt{www.cnet.com} obfuscate advertising URLs when they detect that the ad-blocker is on.
As a result, the resources are translated to the local servers and eventually displayed on the page.
In all of the above instances, performance analysis of ads cannot be achieved through adblocking which limits its scope.

\textbf{\emph{Fine-grained analysis.}} Ad-blockers block content as early as the initiation of network requests, which results in two drawbacks.
First, it prevents fine-grained performance analysis at the browser level.
Second, because the content is blocked at the network request, resources that are further requested by the blocked document during page loading become invisible for inspection.
Furthermore, the current body of works focuses on the network data traffic overhead, neglecting the in-browser computation overhead of ads.

Our approach addresses the above limitations and enables an in-depth performance analysis of ads without adding significant overhead or causing site breakage.
As a result, ad-related activities such as content parsing, rendering, and loading of descendant resources are now visible for analysis.
We present our findings from fine-grained performance characterization in Section \ref{sec:results}.

\subsection{Performance analysis of browsers}
Another notable line of research concerns the performance analysis of browsers given its complexity, large codebase, and multi-process execution.
The majority of browser vendors have an integrated profiler.
Examples include the Chrome profiler~\cite{Chromeprofiler} for Google Chrome and Gecko profiler~\cite{Gecko-profiler} for Mozilla Firefox, which provides statistics about task timing, call graph, memory usage, and network activities.
There have also been several efforts on critical path analysis~\cite{wprofm, wprof, browsers-slow-smartphones, golestani}.
Wprof extracts the dependency graph and breaks down the activities based on type (computation and network activities)~\cite{wprofm, wprof}.
Coz+~\cite{Pourghassemi:2019aa} generates quantitative what-if graphs about the dynamic behavior of the critical path, based on the idea of causal profiling~\cite{cozpaper}.
By analyzing Alexa top webpages, the above studies conclude that computation activities contribute more to the page loading time than network activities.

Our approach to performance analysis is similar in spirit to the above studies.
However, prior efforts did not distinguish between the time spent in the different browser stages/activities based on the resource (ads vs main content).
Since we are mainly interested in analyzing the performance of ads, we address this challenge of distinguishing resources by type and mapping browser activities to resource type for fine-grained performance analysis in this paper.
\section{Browser architecture}
\label{sec:background}
Our performance characterization is based on distinguishing between the amount of work the browser spends on loading the primary or main content of the page and the extra work on loading ad contents.
Discriminating between these two workloads (main vs ad content) requires an understanding of the way browsers load webpages.
Here we outline the browser's high-level design and workflow.

\begin{figure} [htbp]
\centering
\includegraphics[width=0.95\linewidth]{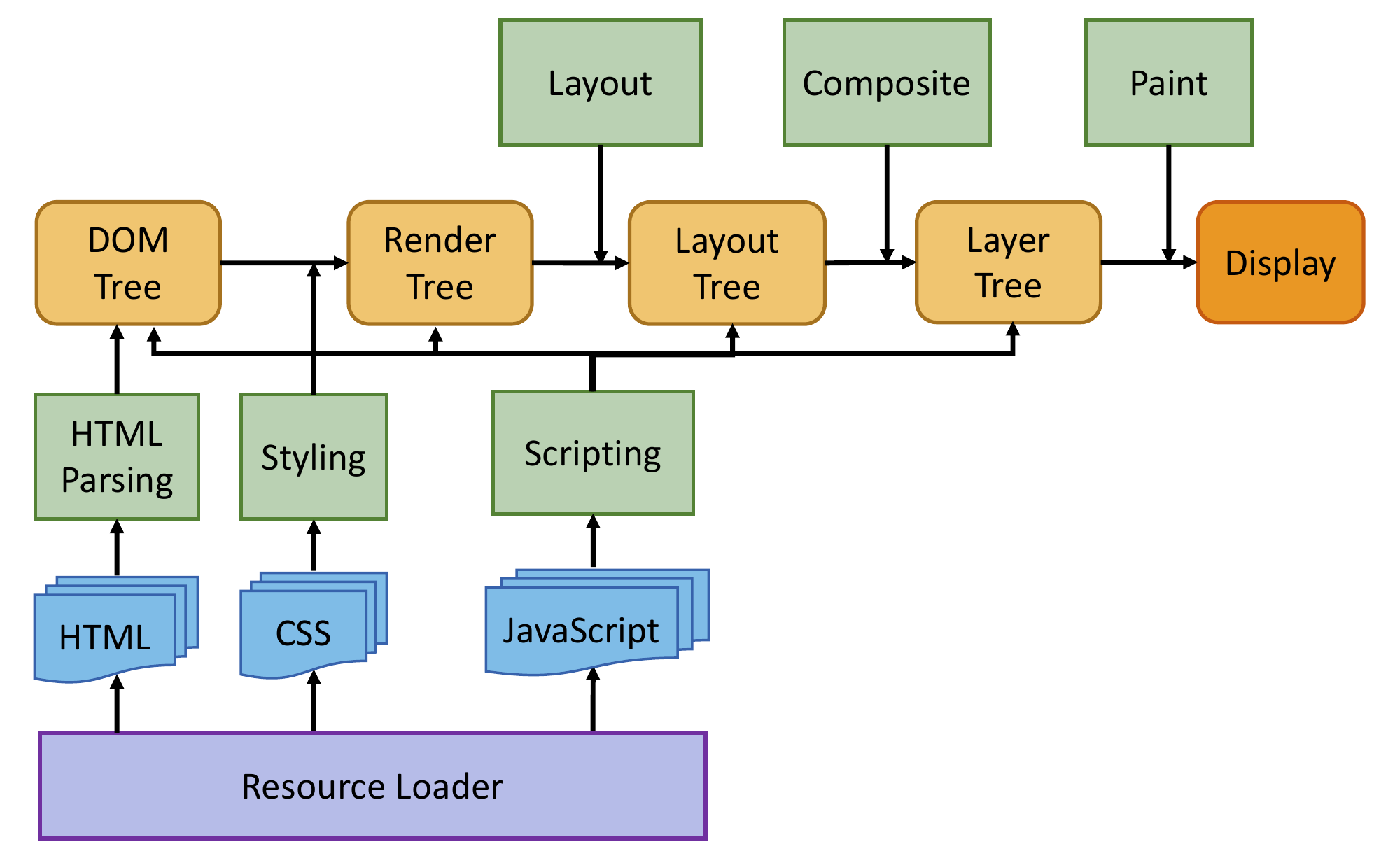}
\caption{High-level architecture of the web browser. The components include the resource loader (purple), six major computation stages (green), and the intermediate trees (yellow) in the page loading pipeline.}
\label{fig:browser_arch}
\end{figure}

Modern browsers have different features and user interfaces but they essentially employ the same architecture to load webpages.
Figure \ref{fig:browser_arch} shows the browser's high-level page loading workflow.
The process begins when the user submits a URL request to the browser interface.
The \emph{resource loader} is responsible for initiating HTTP requests and fetching resources (network activities) from the server.
Once the resource is downloaded (incrementally or fully), the six major computation stages (shown by green color in the figure) evaluate the resources and render the website.
The computation stages are \emph{HTML parsing} (building DOM), \emph{Styling} (evaluating stylesheets and adding attributes to DOM tree), \emph{Scripting} (responding to user interactions and dynamic behavior of the page), \emph{Layout} (evaluating size and position of DOM elements), \emph{Composite} (combining graphical layers), and \emph{Paint} (mapping layers to pixels).
Each of these stages might contain multiple sub-tasks or simply referred to as \emph{activities}. For example, HTML parsing consists of byte stream decoding and preprocessing, tokenizing, and DOM tree construction activities.

\begin{figure} [htbp]
\centering
\includegraphics[width=0.98\linewidth]{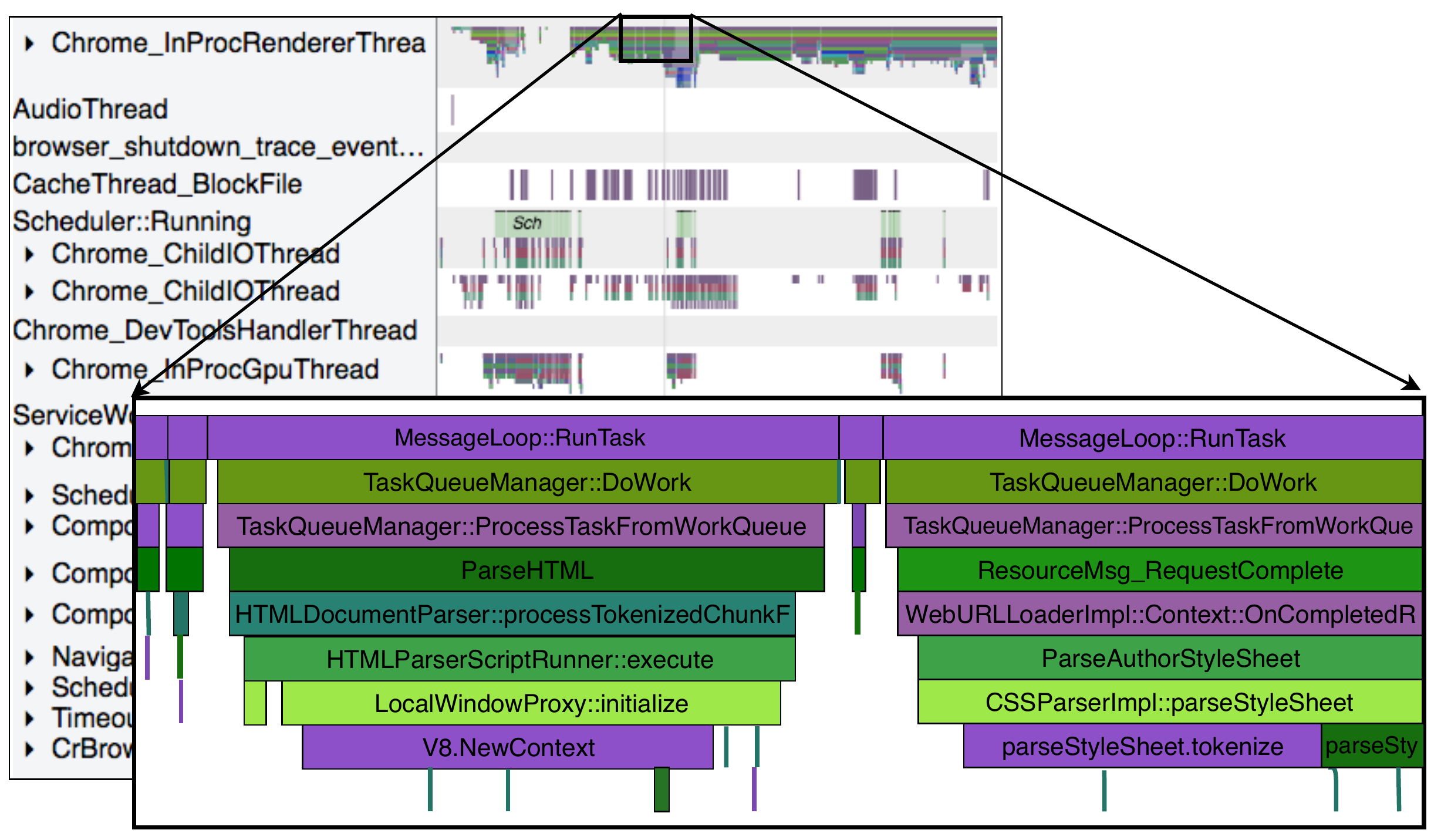}
\caption{Snapshot of the browser activities in loading \texttt{www.apple.com}.}
\label{fig:browser_activities}
\end{figure}

The computation activities are frequently invoked by the browser during page loading.
Figure \ref{fig:browser_activities} illustrates a snapshot of browser activities when loading \texttt{www.apple.com}.
As we can observe, there are numerous instances of such activities during page load.
The order in which these activities are executed is based on the dependency imposed by the page content.
For instance, if JavaScript modifies an attribute of a DOM element, this forces the browser to recalculate the style, update layouts, composite layers, and repaint the screen.
This is commonly known as \emph{reflow} which can have a significant impact on performance.
However, if the JavaScript only modifies the color of a DOM node, the reflow pipeline bypasses layout and if the change does not modify the graphical layers, the reflow also bypasses compositing.
On top of that, browsers exploit parallelization between independent activities to accelerate page load time as shown in the figure.
Due to the dependency between activities, dynamic parallelization, and stochastic behavior of the browser in resource downloading and dispatching tasks, the time of each activity is indeterminate.
Therefore, tracking the dependency chain between browser activities and attributing activities to the corresponding workload type, i.e. ads or non-ads, are challenging tasks.

\section{Methodology and \lowercase{ad}P\lowercase{erf} Infrastructure}
\label{sec:adperf}
To distinguish between the performance cost of web ads from the main content (non-ads), we apply a systematic approach as follows.
First, we \emph{extract} all browser activities that are associated with the page loading process.
Second, we \emph{identify} which resource, i.e. a web document, explicitly or implicitly initiates each browser activity.
Third, we \emph{classify} activities into ads and main content considering the type of the resource corresponding to each activity.
Finally, we \emph{measure} the total execution time spent on each class of activity as a performance index distinguishing the workload in each class. 

\begin{figure} [htbp]
\centering
\includegraphics[width=0.99\linewidth]{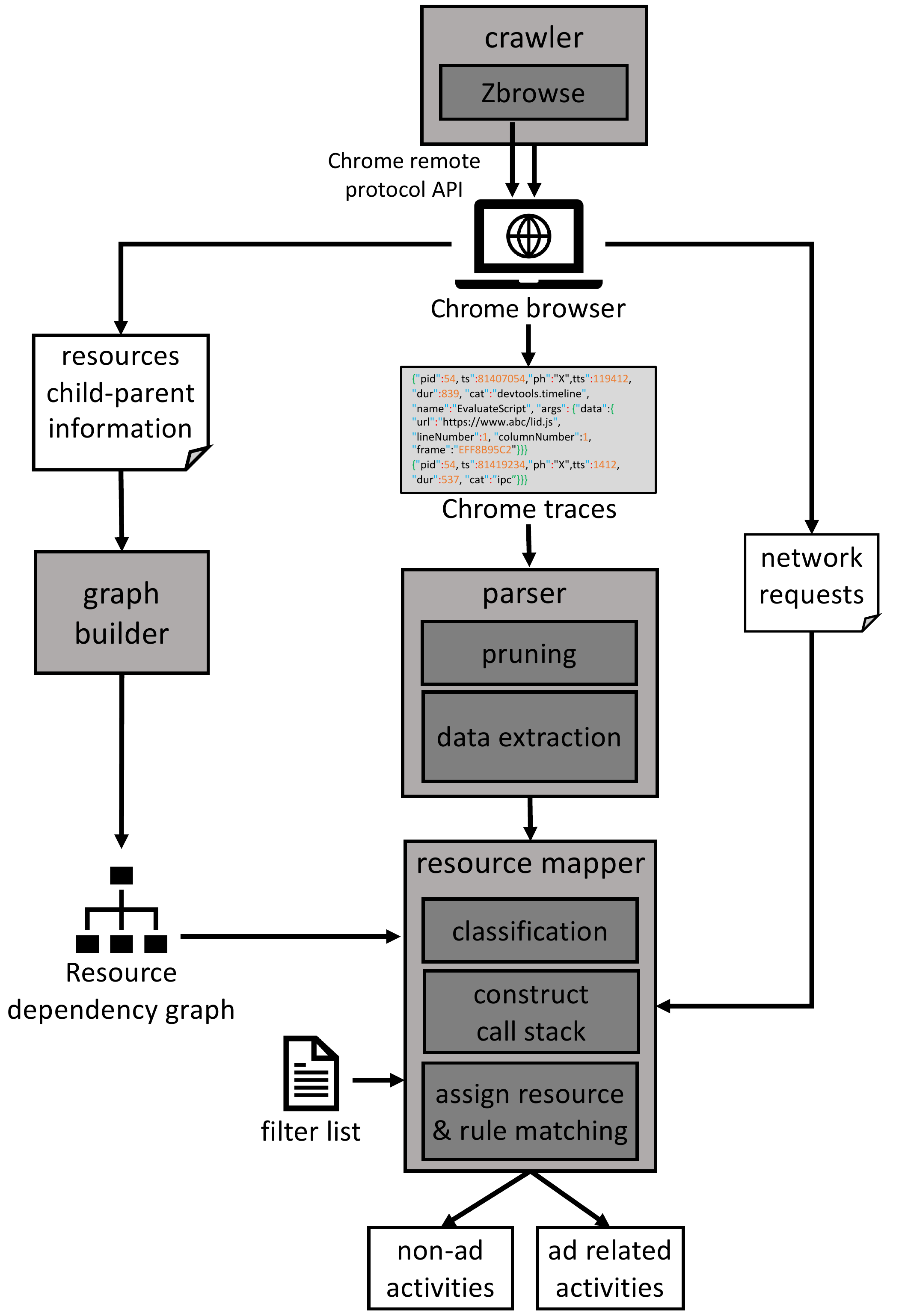}
\caption{Structure of adPerf. AdPerf contains crawler, parser, resource mapper, and graph builder modules that are shown with dark boxes.}
\label{fig:performance_analyzer}
\end{figure}

To realize the above methodology, we implement an infrastructure, named \emph{adPerf}, for the Chrome browser.
Note that adPerf can be extended to other browsers as well since the same technique applies to all browser architectures.
Figure \ref{fig:performance_analyzer} shows the design of adPerf. Below, we describe each module in detail.

\subsection{Crawler}
The first module in adPerf (top of the figure) is a \emph{crawler} (\texttt{Node.js} script) that is responsible for setting up the headless Chrome and crawling websites.
The crawler uses the Chrome remote protocol APIs \cite{protocol} under the hood to interact with the browser and streams Chrome traces \cite{Chromeprofiler} to the file.
Chrome traces are primarily used for profiling and debugging the Chrome browser, so they are designed to be low-overhead.
Tracing macros cost a few thousand clocks at most \cite{Chromeprofiler} and the logging to file happens after the page is loaded.
Chrome traces are capable of recording intermediate browser computing activities including page loading activities in the Blink rendering engine and V8 JavaScript engine with microsecond precision.
Each trace contains some information about the associated activity such as thread id and function arguments.
Below is an example trace for one \emph{Scripting} activity:
\vspace{5pt}
\begin{lstlisting}[language=json,firstnumber=1]
{"pid":54,
 "tid":35,
 "ts":81407054,
 "ph":"X",
 "tts":119412,
 "dur":839,
 "cat":"devtools.timeline",
 "name":"EvaluateScript",
 "args":{"data":{
       "url":"https://www.google-analytics.com/linkid.js", 
       "lineNumber":1,
       "columnNumber":1,
       "frame":"EFF8B95C2"}}}
\end{lstlisting}
\vspace{5pt}

Additionally, the crawler intercepts network requests, i.e. \\ \texttt{onBeforeRequest} event, and extracts the header and body of every HTTP request.
This information will be used later for resource matching.

\subsection{Parser}
When the website is loaded, the raw Chrome traces are fed to the \emph{parser} as shown in the figure.
The adPerf parser does two major tasks.

\textbf{\emph{Pruning.}} It goes through the traces and extracts all page loading activities and prunes the browser-dependent ones (such as browser garbage collection and inter-process communication activities).
Essentially the collected activities are affiliated with one of the six browser stages shown in Figure \ref{fig:browser_arch}.
For instance, the parser considers every trace connected to script evaluation, V8 script compiling, V8 execution, callback functions triggered by browser events or timeout among others as part of the Scripting stage.

\textbf{\emph{Data extraction.}} For each activity, the parser extracts the following data: start time, end time, relative stage, thread and process id, and function arguments if they contain resource information.
This data is necessary to construct the call stack and attribute activities to resources.

\subsection{Resource mapper}
Once the traces are parsed and categorized, this data along with previously obtained network information is fed to the \emph{resource mapper} module.
The task of the resource mapper is to assign each activity to an associated resource.
Unfortunately, we observed that a significant number of traces do not contain any resource information.
Therefore, a key challenge for the resource mapper is to extract this relation.

\begin{figure} [htbp]
\centering
\includegraphics[width=0.98\linewidth]{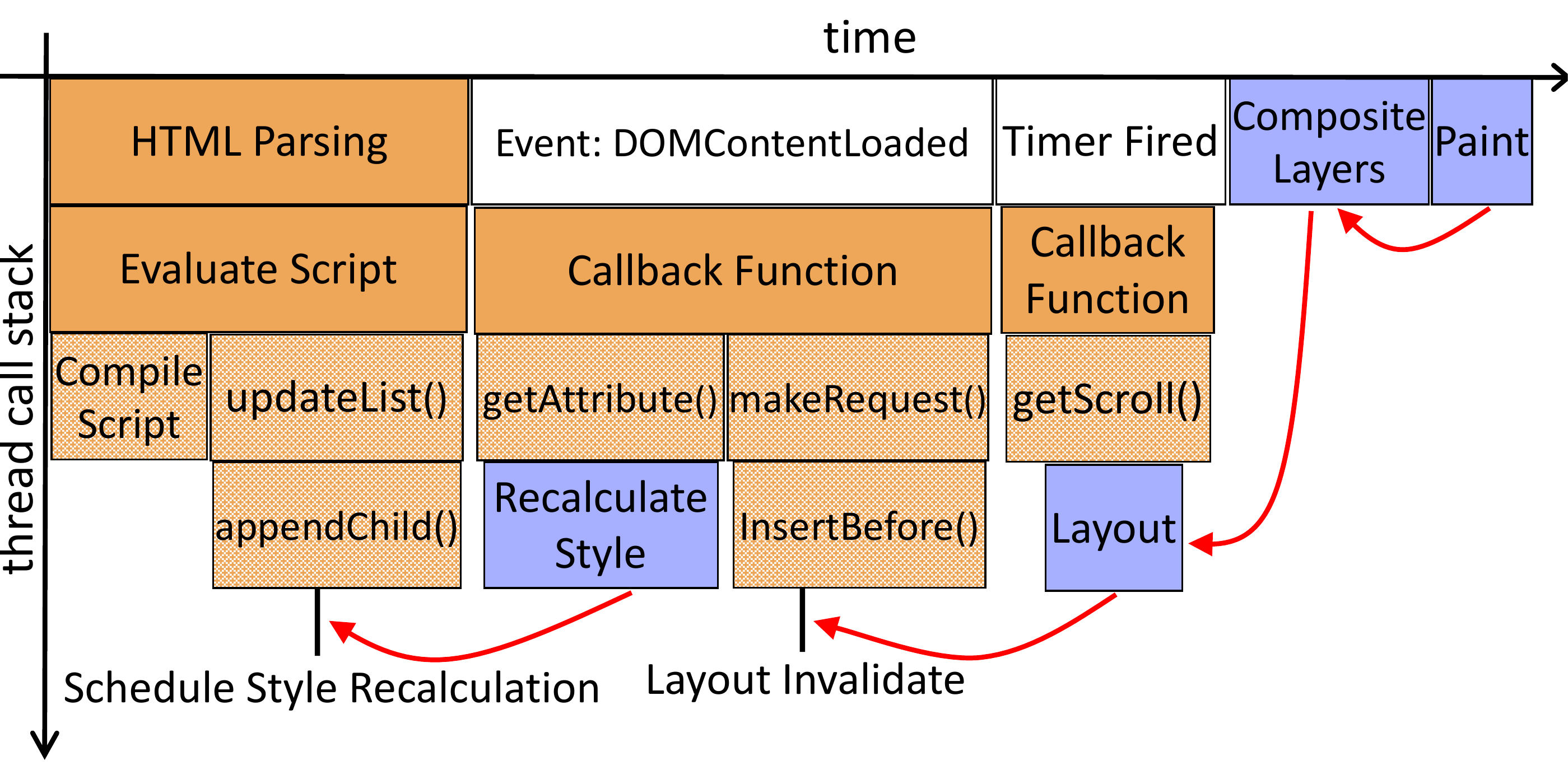}
\caption{Call stack timeline for a Chrome thread constructed by adPerf resource mapper. Resource mapper assigns resource to each activity using information in the traces (orange activities with solid texture) and call stack (orange activities with dotted texture) for parsing and evaluation activities and tracks initiator for tree manipulation and rendering activities (purple activities). }
\label{fig:stack}
\end{figure}

To address this challenge, the resource mapper first builds a call stack of activities for every thread by tracking the start time and end time of activities executed by each thread.
Figure \ref{fig:stack} demonstrates an example activity call stack timeline for a browser thread where activities are shown with boxes.

After constructing call stacks, the resource mapper classifies activities into two groups -- \emph{parsing and evaluation} and \emph{tree manipulation and rendering}.
The former contains activities that \emph{explicitly} relate to a resource such as HTML parsing, image decoding, stylesheet parsing, and JavaScript evaluation which directly operate on a document.
Activities belonging to this group are colored orange in the figure.
The latter contains activities that \emph{implicitly} relate to a resource.
These include activities in styling (except stylesheet parsing which belong to the former group), layout, composite, and paint stages that deal with the browser's intermediate data structures (trees) and display.
Purple activities in the example belong to this group.
Finally, the resource mapper finds the corresponding resource for each activity group as follows.

\textbf{\emph{Parsing and evaluation.}} For the majority of the activities in this group, the resource mapper extracts the resource file information from function parameters extracted by the parser.
Orange activities with solid texture such as HTML Parsing and Callback Function in Figure \ref{fig:stack} are examples of activities that the document on which they parse or evaluate can be determined from frame id and resource file data in their traces.
However, a small number of activities do not contain any resource information.
For activities with unresolved resource files (activities shown with an orange color and dotted texture in the figure), the resource mapper uses the constructed call stack and follows their ancestors and associates them with the caller's resource file.
For example, \emph{appendChild} JavaScript function is called by \emph{updateList} and this function along with \emph{Compile Script} activity are invoked by \emph{Evaluate Script} activity that is previously assigned to a JavaScript document.

\textbf{\emph{Tree manipulation and rendering.}} For this group, we have to distinguish between the different resources that implicitly trigger the activities that belong to this group.
For styling activities, we observe that Chrome recalculates styles after the \emph{Schedule Style Recalculation} event is fired.
As seen from Figure \ref{fig:stack}, this event is fired in the middle of \emph{parsing and evaluation} of a resource (typically JavaScript document) that attempts to modify the DOM node style.
We track the call stack for this event to the initiated \emph{parsing and evaluation activity} and relate this styling activity to the triggered document.
Similarly, for layout, Chrome updates layout tree when \emph{Layout Invalidate} event is fired.
In our example, this is fired when the command \emph{this.\_util.elem.innerHTML=e} is executed in the \emph{InsertBefore()} function.
We use a similar procedure as styling to relate layout activities to the initiating resource from the call stack of the \emph{Layout Invalidate} event.

Note that the browser does not necessarily update the style and layout of nodes immediately after the events are triggered. 
Depending on the priority of other activities in the task scheduler queue, the browser might dispatch these activities later.
As a result, when a resource triggers one of these two events (\emph{Schedule Style Recalculation} or \emph{Layout Invalidate}), a second resource may fire one of these two events again before the browser updates the tree.
In this case, we consider the first resource as the initiator since the tree will be traversed and updated even in the absence of the second resource.
Chrome tends to composite and/or paint immediately after styling or layout which leads to repaint.
Therefore, the associated resource for the composite and paint activities simply derives by following the chain to the last executed styling or layout activity as shown by the red arrows in the figure.

Once page loading activities have been assigned to the resources, adPerf uses network data from the crawler to link the resources to the associated network requests (i.e. URLs).
Then it uses a filter list to distinguish between ad resources and non-ad resources.
We use EasyList \cite{easylist}, the primary and most popular filter rules list for advertisements, for our experiments.
However, users can also provide their own custom filter rules.
adPerf employs adblockparser (an optimized python package \cite{adblockparser}) to match the URLs against filter rules.
One might think that since our methodology uses an identical rule matching procedure to ad-blockers, it might incur a similar overhead.
This is however not the case since that rule matching in adPerf is passive and does not steal computation cycles from the page loading process.
Finally, adPerf reports the execution time of the page loading activities categorized by ads and non-ads.

\subsection{Graph builder}
\label{subsec:graph}

There exist dependencies between resources on the website.
For instance, let's say a website downloads a JavaScript file from a third-party domain.
In this file, it can further request an image or an HTML document from another domain and this chain can go deeper.
To evaluate the performance cost of different sources such as ad domains and to further evaluate their trustworthiness (Section~\ref{sec:results}) requires first tracing this resource dependency chain and building a \emph{resource dependency graph}.

We extract the dependency between resources of the websites using Zbrowse \cite{zbrowse}.
Zbrowse exploits Chrome devTools protocol that allows us to instrument, inspect, and debug the Chrome browser.
It also generates the child-parent relation for every network request.
We embed Zbrowse in the adPerf crawler module as shown in Figure \ref{fig:performance_analyzer}.
This way, we can extract the resources child-parent data at the same time when we crawl the websites.
The \emph{graph builder} module uses Zbrowse's output and constructs the dependency graph for resources. 
In cases where third-party JavaScript gets loaded into a first-party context and makes an AJAX request, the HTTP referrer appears to be the first-party. 
We follow \cite{ikram2019chain} and allow the graph builder to conserve this relation and include the URL of the third-party from which the JavaScript was loaded.
Since one resource can, in turn, request multiple resources, the constructed graph has the shape of a tree rather than simple chains of dependencies.

\begin{figure}[htbp]
\centering
\includegraphics[width=0.95\linewidth]{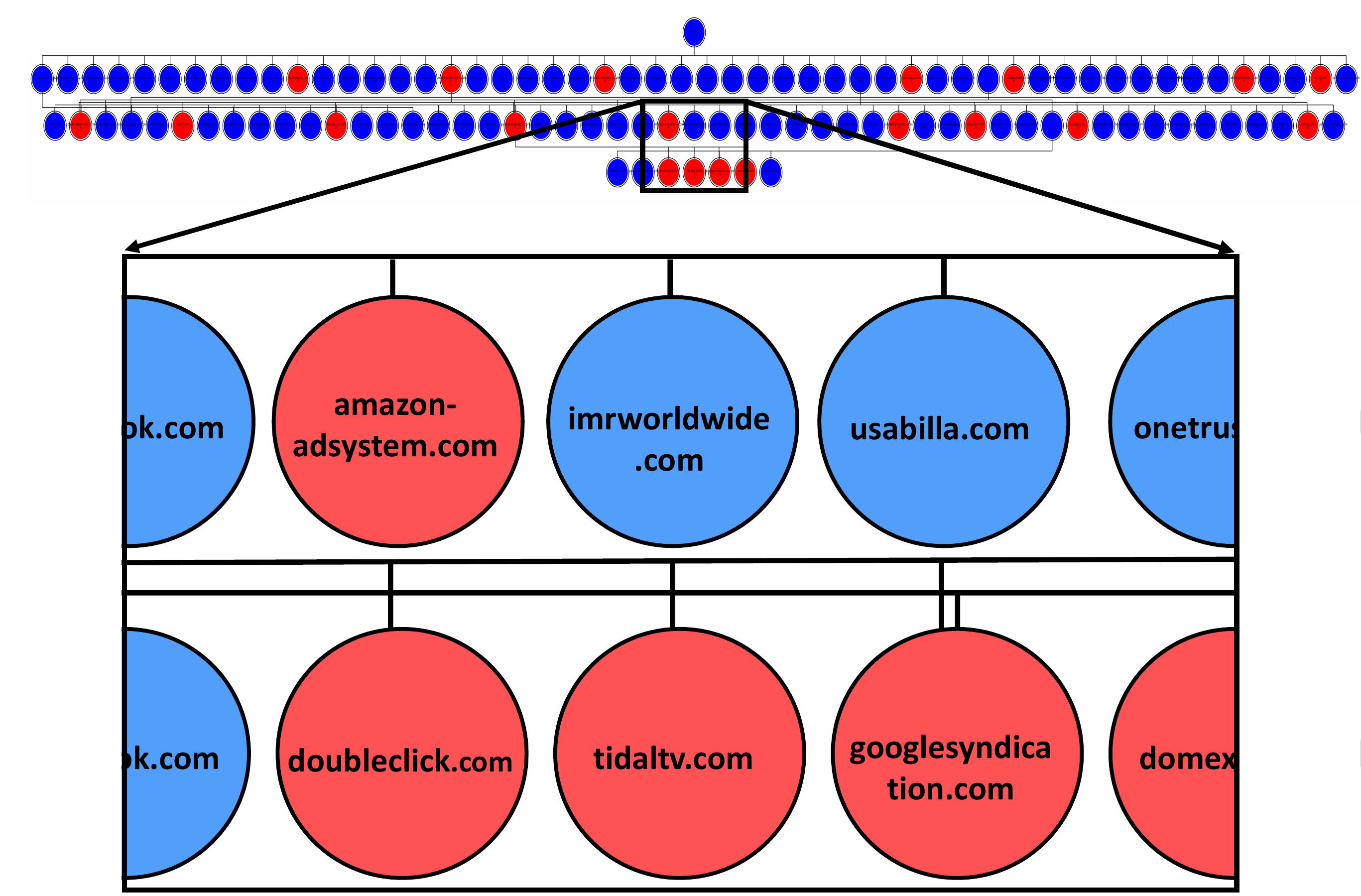}
\caption{Resource-dependency graph for \texttt{www.cnn.com}. Ad nodes are colored red and non-ad nodes are colored blue.}
\label{fig:dependency_graph}
\end{figure}

Figure \ref{fig:dependency_graph} shows this graph for an example website, \texttt{www.cnn.com}.
Here, we combine the resources from the same domain (at each level) into one node for easier visualization.
The root node is the publisher and the remaining nodes are referred to as third-party domains.
For differentiation, we color ad nodes (domains that deliver at least one ad resource) red and non-ad nodes (domains without any ad resources) blue in this graph.
As we can see from the figure, a considerable number of third-party domains are ad nodes. 
This is a concerning finding since typically publishers are not aware of the contents delivered by third-party websites. 
Generally, publishers trust the first-party domains (in the first-level of tree) but those websites might deliver their contents from another website or chain of websites that are not verified by the publishers.
\section{Experimental setup}
\label{sec:setup}
\emph{\textbf{System.}} The experiments are conducted on a MacBook Pro with 2 cores and 8 GB RAM.

\noindent \emph{\textbf{Network.}} The system is connected to the WiFi with a 400 Mbps network connection. 
To obtain accurate result about communication overhead, we do not set up any proxy and/or local server.

\noindent \emph{\textbf{Test corpus.}} Our test corpus consists of two sets of web pages -- (a) top 350 websites from Alexa top500 news list \cite{alexa} and (b) top 200 websites from Alexa top500 list \cite{alexaus}.
We will refer to these two web page datasets as \emph{news} and \emph{general} respectively.
For each dataset, we crawl the corresponding corpus twice.
The first time, we crawl the home page or landing page of the website.
The second time, we randomly click a link on the home page and crawl the page that it leads to.
We exploit Chrome Popeteer \cite{popet} to automate link clicking.
In our experiments, the former is referred to as the \emph{landing} page crawl and the latter is referred to as the \emph{post-click} page crawl.

\noindent \emph{\textbf{Experimental repeat.}} In each crawl over the corpus (total 4 crawls), we load websites at least 3 times to account for fluctuations in page loading.

\noindent \emph{\textbf{Evaluation domain.}} Since the main goal is to characterize the performance cost of ads, we primarily provide evaluation results for the websites that contain ads.
This is about 80\% of news websites and 40\% of top general websites.

\section{Results and Discussion}
\label{sec:results}
In this section, we analyze the performance cost of ads both at the level of the ad domains (close to the origin) and deeper in the browser (close to the metal).
First, we analyze the performance cost of ads on the websites using adPerf broken down by costs incurred by the computation (i.e. rendering engine) and network (i.e. resource loader).
Then, we investigate a level deeper to further understand which computation stages and which network resources contribute more to the computation and network ad costs respectively.
Finally, we zoom out and analyze the ad domains themselves to demystify their contribution to the performance cost of web ads.

\subsection{Computation cost of ads}

\begin{figure} [htbp]
\centering
\includegraphics[width=0.95\linewidth]{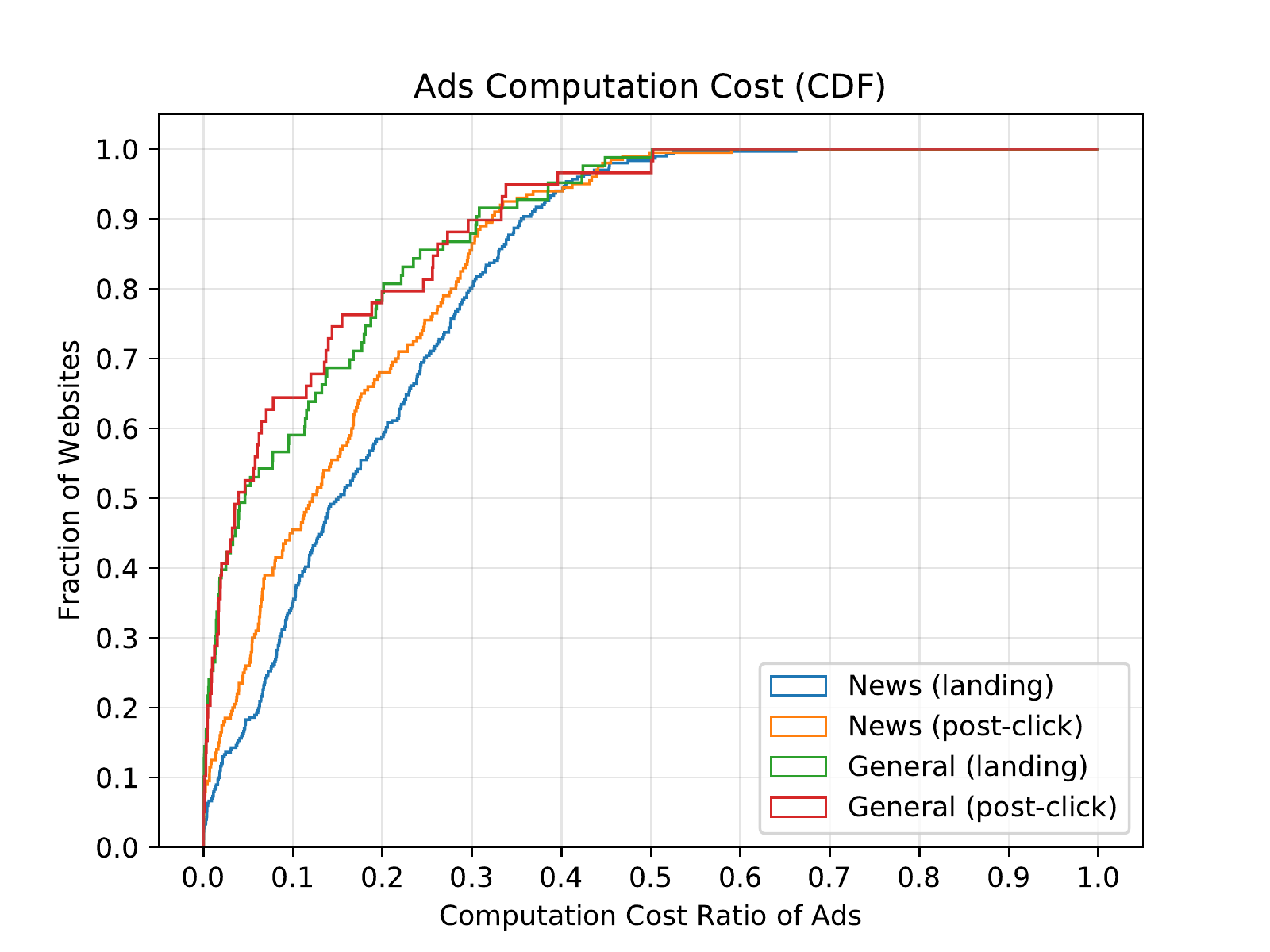}
\caption{Computation cost of ads in two datasets namely top general and top news websites. Each domain in the dataset is crawled twice (landing page and post-click page). }
\label{fig:computation_cost}
\end{figure}

For every website, we calculate the fraction of time spent in ad-related activities to the total activities (ad + non-ad).
Figure \ref{fig:computation_cost} shows the CDF distribution of this fraction for the 4 different crawls.

\textbf{Finding 1.} According to the figure, web ads have a significant impact on the performance of the website.
For example, half of the news websites spend more than 15\% of their computing workload on ads. Interestingly, 20\% of them take more than 30\% of the time on advertising which can be concerning from the user's perspective.
It also motivates website builders and ad providers to optimize their advert contents.
Compared to the news sites, ads have a lower cost on the general corpus.
The median in this corpus is 5\%.

\textbf{Finding 2.} The figure presents another interesting detail when we compare the landing and post-click page graphs.
Ads have a higher performance cost when loading the landing page versus the post-click page of news websites by about 25\% on average.
However, this is not the case for general websites.
Post-click pages of popular general websites have almost similar cost-performant ads as the landing page.
Further, we aggregate the total time spent on ad-activities across all browser stages and compare that to the total time spent on the main content.
The average percentage of time spent on ads versus main content for the news landing page, news post-click, general landing page, and general post-click datasets is 17, 15, 11, and 10\% respectively.
The averages are higher than the median percentages reported earlier because a small number of websites spend 40-50\% of the computation time on ad-activities. 

\begin{figure} [htbp]
\centering
\includegraphics[width=0.95\linewidth]{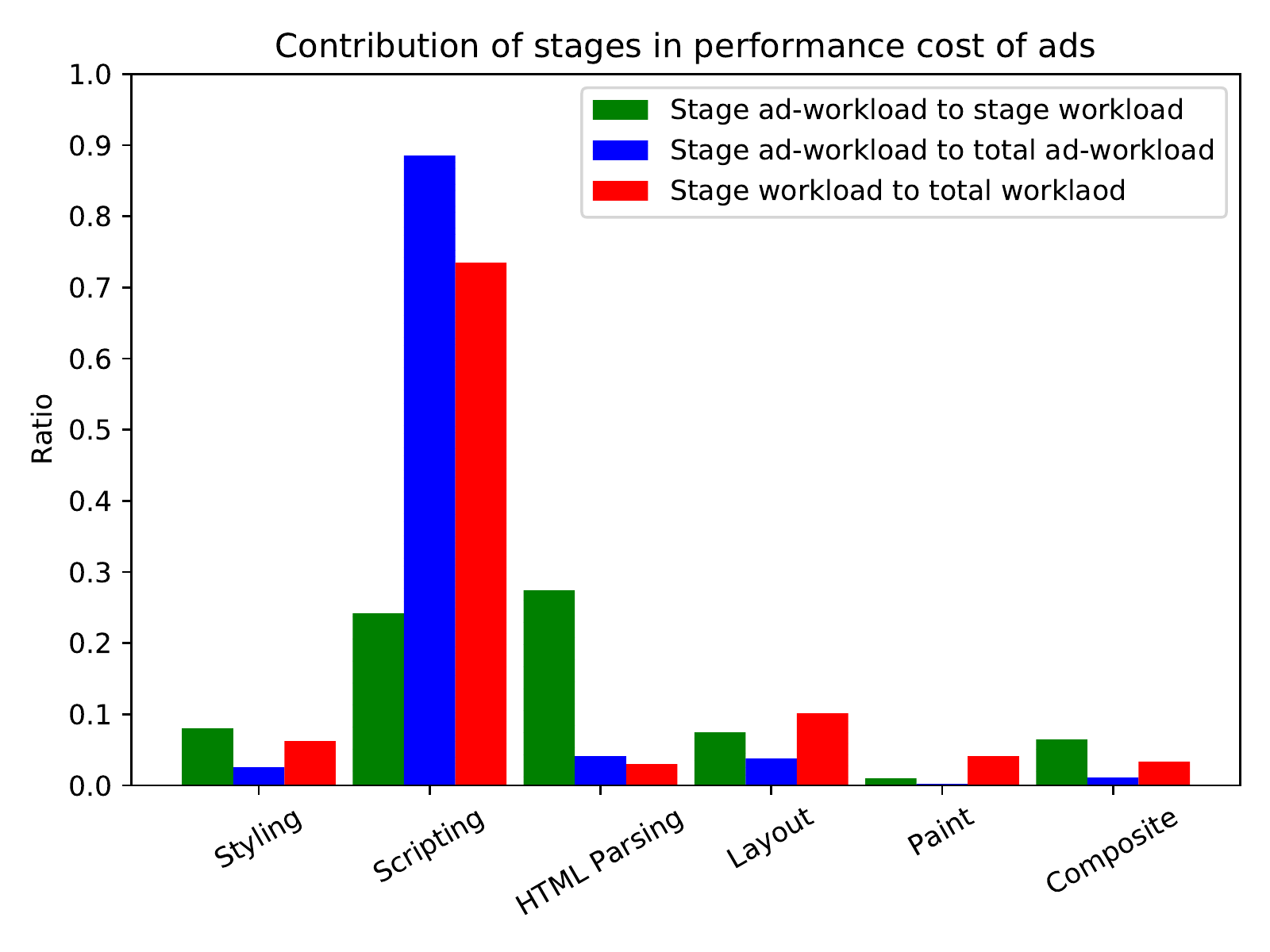}
\caption{Contribution of the different browser stages to the performance cost of ads for the news landing corpus. The three bars for each stage correspond to the three ratio metrics ($ct_{ad}^{s}/ct_{*}^{s}$, $ct_{ad}^{s}/ct_{ad}^{*}$, and $ct_{*}^{s}/ct_{*}^{*}$ from left to right).}
\label{fig:stage_ratio}
\end{figure}

\emph{\textbf{Breakdown of ad computation cost.}} Since we observe that ads can have a significant impact on website loading, it is worthwhile to explore where this overhead comes from.
This will enable website builders and ad providers to focus their optimization efforts on those activities that are the primary sources of performance loss.
Accordingly, we classify the computation cost of ads by the granularity of the browser stages (outlined in Section~\ref{sec:background}).
Figure \ref{fig:stage_ratio} shows the contribution of the six major stages for the news corpus.
For each stage, $s$, we measure the following three metrics.
Note that $ct^{s}$ is the computation time of stage $s$ while $ct^{*}$ is the total time spent in computation across all the browser stages.
Similarity, $ct_{ad}$ is the computation time spent on ad-activities while $ct_{*}$ is the total time spent on all activities.
Therefore, $ct_{*}^{*}$ is the total time of all computation activities in the browser.

\begin{enumerate}
\item The fraction of time spent on ad-activities in stage $s$ to the total time spent on all activities in stage $s$ [$ct_{ad}^{s}/ct_{*}^{s}$].
This is shown by the green bars.
This is an intra-stage metric depicting how the stage workload is split between ads and the main content.

\item The fraction of time spent on ad-activities in stage $s$ to the total time spent on ad-activities across all stages [$ct_{ad}^{s}/ct_{ad}^{*}$].
This is shown by the blue bars.
This is an inter-stage metric showing how much one stage deals with ads compared to the other stages.

\item The fraction of time spent on all activities in stage $s$ to the total page load computation time or total workload [$ct_{*}^{s}/ct_{*}^{*}$].
This is shown by the red bars.
This is another inter-stage metric but unlike the second metric, it shows the influence of a particular stage, $s$ on the entire page load.
\end{enumerate}

It is important to correlate both the inter-stage metrics to have a complete analysis.
For example, if a stage shows a considerable contribution to ads (i.e. second metric) but has very little impact on page loading (i.e. third metric), that stage likely has a moderate impact on the performance optimization of ads.

\textbf{Finding 3.} As we can observe from Figure \ref{fig:stage_ratio}, \emph{scripting} has the highest impact, more than 88\%, on the computation cost of ads.
Incidentally, it also has a significant impact (73\%) on the computation workload of the entire page load.
The difference between these two metrics indicates that ads are more scripting heavy than the total workload.
This is because ad-content shows 21\% more dynamic characteristics than the original page content in our news corpus which increases the time spent in the scripting stage.
At the same time, this stage only spends 25\% of its time on ad-related content (i.e. first metric) which is an interesting observation since it reveals that ads are not the primary bottleneck of the scripting stage but improving this stage will considerably improve the performance of ads as scripting is the major workload of today's web ads on news sites. 

\textbf{Finding 4.} Another observation from Figure \ref{fig:stage_ratio} is that HTML parsing has a minor influence on page loading, i.e less than 5\% in comparison with scripting but ads have more impact on this stage (comparing green bars).
In other words, optimizing ads HTML code is expected to improve HTML parsing workload more than optimizing ads JavaScripts for the scripting stage, even though HTML optimizations can only marginally improve page load time.
This underscores the importance of correlating the intra- and inter-stage metrics to accurately guide optimization efforts.
We observe similar behavior for the general corpus as well.

\subsection{Network cost of ads}

\begin{figure} [htbp]
\centering
\includegraphics[width=0.95\linewidth]{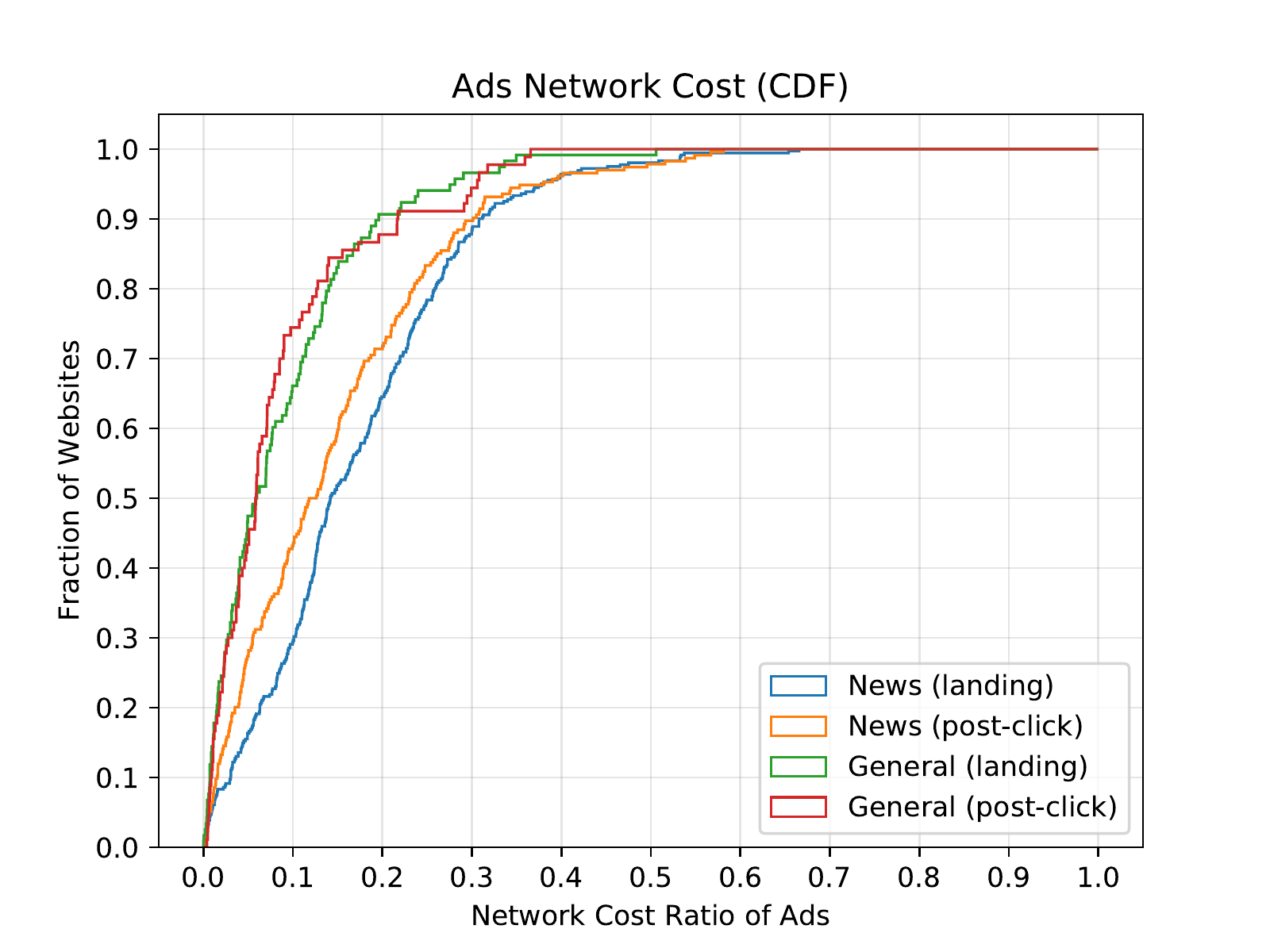}
\caption{Network performance cost of ads over websites in two corpus's: top general and news websites. Each corpus contains landing and post-click landing pages.}
\label{fig:network_cost}
\end{figure}

Beside computation activities, loading ads imposes overhead on the network activities.
To measure the performance cost of ads over the network, for each website, we calculate the ratio of time spent on fetching ad-related resources to the total time spent on fetching all the requested resources.
Figure \ref{fig:network_cost} shows the CDF of this network cost ratio for the $4$ crawls.

\textbf{Finding 5.} The four distributions follow the same order as in Figure \ref{fig:computation_cost} (computation cost of ads), i.e. news websites incur higher network performance cost than general websites.
This is not surprising since more and/or larger ad resources also require more work in parsing, evaluating, and rendering.
According to the figure, the median of the network-cost ratio is 15\% for news websites' landing page and 3\% less on the post-click page.
For the general websites, the median is 6\% for the landing page and post-click page respectively.

\def\arraystretch{1.5}
\begin{table*}[htbp]
\centering
\caption{Summary of the three metrics each for the number of resources and network time spent on resources across two types of pages (landing page denoted by L and post-click page denoted by PC) for the news corpus.}
\begin{tabular}{|c | c | c | c | c | c | c | c | c | c | c | c | c |}
\toprule
Content type ($c$) & \multicolumn{6}{c|}{Stats for the number of resources ($nr$)} & \multicolumn{6}{c|}{Stats for the network time spent on resources ($nt$) }\\
\midrule
\hline
& \multicolumn{2}{c|}{$nr^{c}_{ad}/nr^{c}_{*}$} & \multicolumn{2}{c|}{$nr^{c}_{ad}/nr^{*}_{ad}$} & \multicolumn{2}{c|}{$nr^{c}_{*}/nr^{*}_{*}$} & \multicolumn{2}{c|}{$nt_{ad}^{c}/nt_{*}^{c}$} & \multicolumn{2}{c|}{$nt_{ad}^{c}/nt_{ad}^{*}$} & \multicolumn{2}{c|}{$nt_{*}^{c}/nt_{*}^{*}$} \\
\hline
& L & PC & L & PC & L & PC & L & PC & L & PC & L & PC \\
\hline
Script & 0.235 & 0.218 & \bf{0.415} & \bf{0.449} & \bf{0.398} & \bf{0.431} & 0.251 & 0.236 & \bf{0.489} & \bf{0.572} & 0.326 & 0.369 \\
\hline
HTML & \bf{0.357} & \bf{0.34} & 0.087 & 0.095 & 0.055 & 0.059 & \bf{0.169} & \bf{0.136} & 0.044 & 0.047 & 0.043 & 0.053 \\
\hline
Image & 0.228 & 0.221 & \bf{0.371} & \bf{0.352} & \bf{0.367} & \bf{0.332} & 0.13 & 0.117 & \bf{0.392} & \bf{0.325} & 0.505 & 0.425 \\
\hline
Font & 0.129 & 0.06 & 0.013 & 0.007 & 0.023 & 0.024 & 0.065 & 0.028 & 0.006 & 0.004 & 0.016 & 0.02 \\
\hline
CSS & 0.058 & 0.028 & 0.013 & 0.008 & 0.051 & 0.061 & 0.049 & 0.019 & 0.009 & 0.005 & 0.03 & 0.039 \\
\hline
XML & \bf{0.542} & \bf{0.457} & 0.006 & 0.003 & 0.002 & 0.001 & \bf{0.681} & \bf{0.426} & \bf{0.008} & \bf{0.003} & 0.002 & 0.001 \\
\hline
XHR & 0.179 & 0.125 & 0.045 & 0.032 & 0.057 & 0.053 & 0.122 & 0.069 & \bf{0.048} & \bf{0.038} & \bf{0.066} & \bf{0.085} \\
\hline
Media & 0.042 & 0.044 & <0.001 & <0.001 & 0.002 & 0.002 & 0.034 & 0.029 & <0.001 & <0.001 & 0.001 & 0.001 \\
\hline
Unknown & 0.245 & 0.299 & 0.05 & 0.054 & 0.046 & 0.038 & 0.059 & 0.134 & 0.004 & 0.006 & 0.011 & 0.007 \\
\bottomrule
\end{tabular}
\label{tab:contentType}
\end{table*}

\textbf{\emph{Breakdown of ad network cost.}} To dissect the network costs of ads, we breakdown the network time consumption by content type (such as HTML, image, media, etc).
For each content type, Table \ref{tab:contentType} summarizes statistics about the frequency of resources and network time spent on fetching those resources for the news corpus for both landing and post-click pages.
Given, the number of resources, $nr$ and network time spent on the resources, $nt$, we define three metrics for each (similar to computation stages) as follows.
\begin{itemize}
\item Metrics for the number of resources ($nr$).
\begin{enumerate}
\item The fraction of the number of resources of content type, $c$ to the total number of resources of $c$ [$nr^{c}_{ad}/nr^{c}_{*}$] (intra resource-type metric).
\item The fraction of the number of ad-resources of content type, $c$ to the total number of ad-resources (of all content types)[$nr^{c}_{ad}/nr^{*}_{ad}$].
\item The fraction of the number of resources of content type, $c$ to the total number of resources [$nr^{c}_{*}/nr^{*}_{*}$] to highlight the popularity of the content type.
\end{enumerate}

\item Metrics for the network time spent on resources ($nt$).
\begin{enumerate}
\item The fraction of the network time spent on ad-resources of content type, $c$ to the total network time spent on resources of $c$ [$nt_{ad}^{c}/nt_{*}^{c}$].
\item The fraction of the network time spent on ad-resources of content type, $c$ to the total network time spent on ad-resources (of all content types) [$nt_{ad}^{c}/nt_{ad}^{*}$].
\item The fraction of the network time spent on resources of content type, $c$ to the total network time spent on all resources [$nt_{*}^{c}/nt_{*}^{*}$] to highlight the performance impact of content type, $c$

\end{enumerate}
\end{itemize}

For instance, the first metric for network time for CSS refers to the fraction of time spent on fetching ad-related CSS resources to the time spent on fetching all CSS resources [$nt_{ad}^{css}/nt_{*}^{css}$].

\textbf{Finding 6.} Among all content types, Table~\ref{tab:contentType} shows that XML has the largest percentage of ad resources for both landing (54\% which take up 68\% of the network time in fetching this type of resources from metric 1) and post-click pages (46\% which take up 43\% of the network time).
However, it contributes to an insignificant fraction of the network performance cost for both pages (metric 2).
On the contrary, scripts and images are commonly used by ad providers.
These two types of resources alone makeup nearly 80\% of all ad resources (metric 2) \emph{and} all resources (metric 3) for both landing and post-click pages.
However, comparing metrics 2 and 3 for scripts and images shows that scripts are on average over 20\% more popular than images for post-click pages compared to the landing page.
Script files used in advertising alone are responsible for almost half of the network performance cost of ads, followed by images at ~40\% for landing pages (metric 2).
Since scripts are more popular in post-click pages, they correspondingly also contribute more to the network time spent in ads (57\%) for these pages compared to images (33\%).

\textbf{Finding 7.} Ad-related HTML files constitute 34-36\% of total HTML files but they only take 14-17\% of download time.
A deeper investigation shows that ad HTML documents have a significantly small number of tags (on average $7$) including only one or two <script> tags that encapsulate small and minified code compared to the main-content HTML files with $410$ tags.
This shows that HTML files used for ads are typically lighter than normal HTML files.
Surprisingly, XHR (XMLHttpRequest) resources make up a significant 7\% of the network performance cost for the landing page and 9\% for post-click pages (metric 3).
The corresponding time spent on ad resources is 5\% and 4\% respectively (metric 2).
\subsection{Breakdown of ad performance by source}
\label{sec:results-source}
The results so far breakdown the performance cost of web ads at the lower level of granularity.
Now, we zoom out and attribute the performance cost of ads based on their origin, i.e. ad domains.
This allows us to gain a clear picture of the third-party ad domains and how they contribute to the performance cost.
Accordingly, we build the resource-dependency graph (as described in Section \ref{sec:adperf}) for all news websites in our test corpus.
Overall we identify more than 300 distinct ad domains.

\begin{figure} [htbp]
\centering
\includegraphics[width=0.95\linewidth]{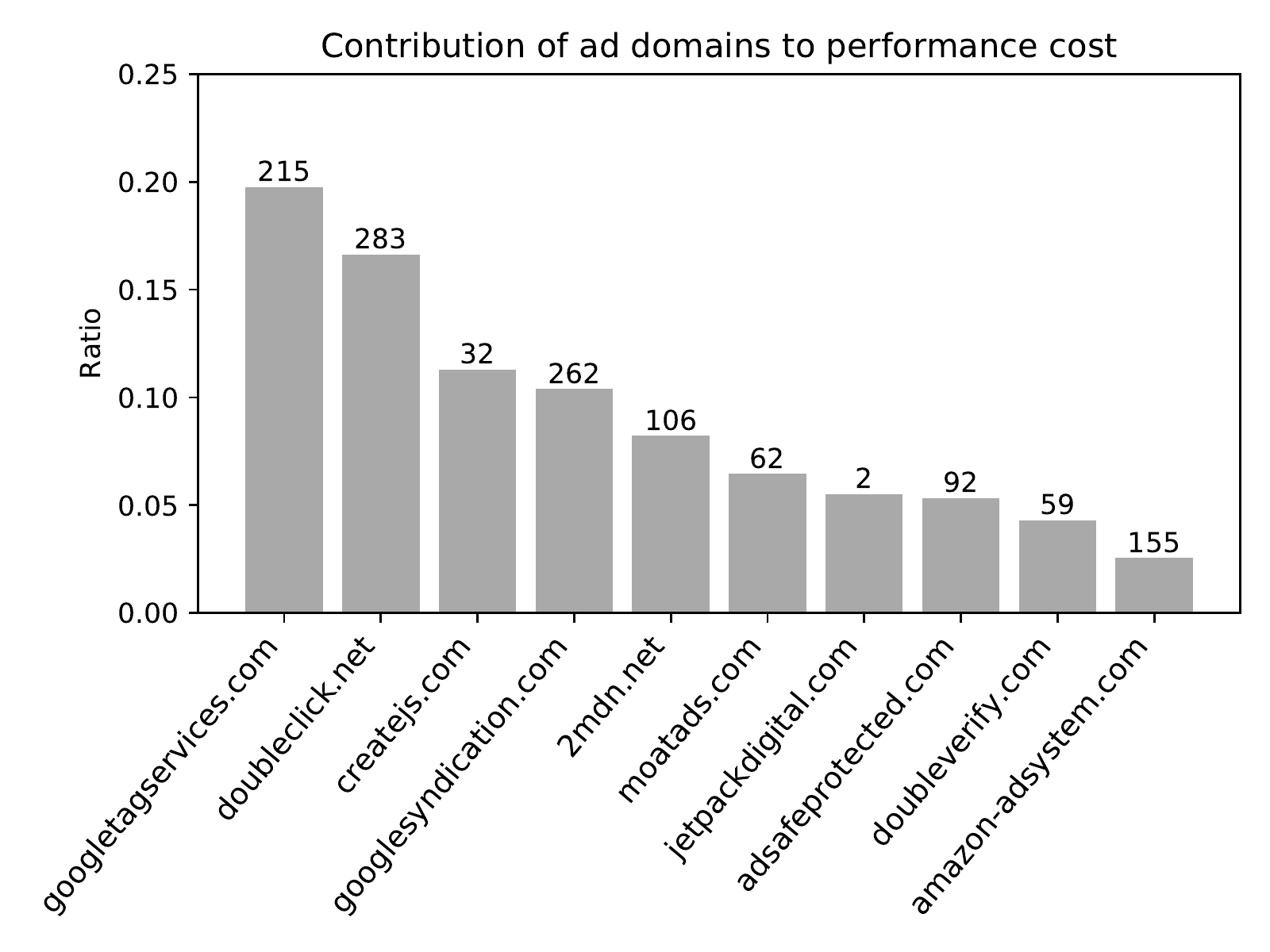}
\caption{Contribution of ad domains to the computation cost of web ads. The number on top of each bar is the number of websites serviced by that particular ad domain.}\label{fig:domain_contribution}
\end{figure}

\textbf{\emph{Breakdown of computation performance cost by ad domains.}} 
For every ad domain, we first aggregate the time the rendering engine spends on evaluating the resources that are delivered by that domain.
We also measure the total time spent on ads through the crawl (ads computation cost).
Then, we calculate the ratio between the above two measures which is an indicator of how third-party ad domains contribute to the computation cost of ads.
Finally, we sort the ad domains by this ratio and plot the contribution of the top 10 domains (out of 300) in Figure \ref{fig:domain_contribution}.
The number on top of each bar is the number of websites in our corpus that are referred to that ad domain.

\textbf{Finding 8.} The largest contribution to the computation of ads on the web are from \texttt{googletagservices.com} and \texttt{doubleclick.net}. 
The former is a Google tag management system for managing JavaScript and HTML tags used for tracking and analytics on websites and the latter is a popular ad provider.
Together, they also deliver about 35\% of the total ad resources.
Our results also show that not all the ads are delivered by well-known ad domains and 50\% of ad domains appear only in the dependency graph of \emph{one} website.

Besides, the data shows that the number of websites serviced by an ad domain is not an indicator of its performance cost.  
For instance, \texttt{googlesyndication.com} has approximately the same contribution to the performance cost of ads as \texttt{createjs.com} but it services over $8 \times$ more websites than the latter. 

\begin{figure} [htbp]
\centering
\includegraphics[width=0.95\linewidth]{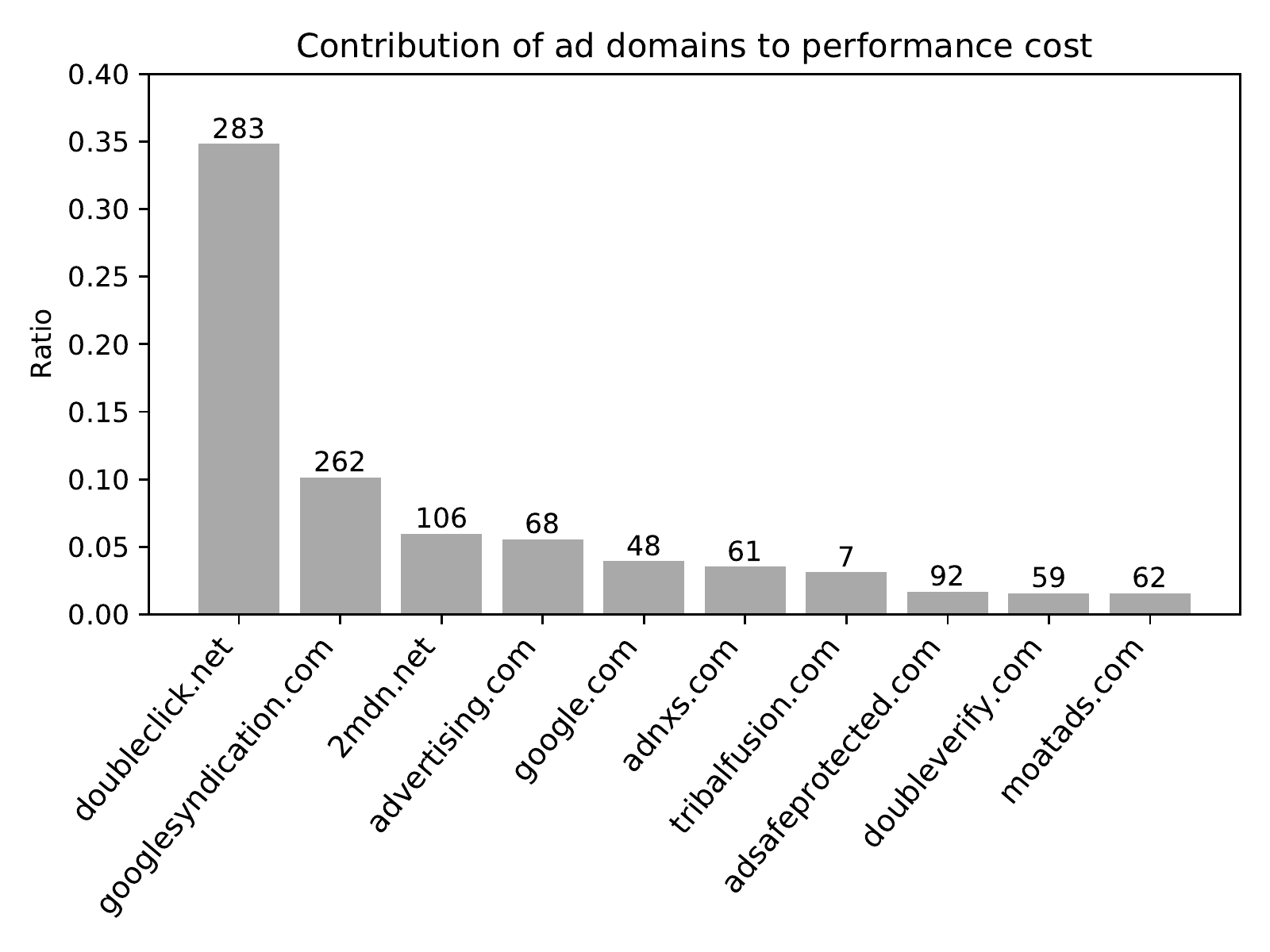}
\caption{Contribution of ad domains to the network cost of web ads. The number on top of each bar is the number of websites serviced by that particular ad domain.}\label{fig:domain_contribution_network}
\vspace{-1em}
\end{figure}

\textbf{\emph{Breakdown of network cost by ad domains.}} We follow a similar procedure as above for estimating the contribution of individual ad domains to the network cost of a page load.
For every ad domain, we first aggregate the time the browser spends on fetching resources by that domain,
Then, we calculate the ratio of the total time spent on fetching ad resources in our crawl to the above time.
Figure \ref{fig:domain_contribution_network} shows the top $10$ ad domains that have the highest contribution to the network cost of ads in the news corpus.

\textbf{Finding 9.} About 35\% of the network cost of web ads on news websites is traced to \texttt{doubleclick.net} followed by the popular ad syndication \texttt{googleadsyndication.com} with 10\% contribution.
These two domains along with other domains maintained by Google constitute approximately 51\% of the total ad network cost.

\textbf{Finding 10.} Contrasting the computation cost of domains with their network cost shows that these two performance costs are correlated.
This is expected since fetching more and larger documents take a longer time to evaluate and display.
Interestingly, we also observe domains that have a high computation cost but insignificant network cost or vice versa.
For instance, \texttt{googletagservices.com} has the \emph{highest} contribution (19.7\%) to the computation cost of ads among all 300 ad domains.
However, it contributes to less than 1\% of the network cost (ranked 16 and therefore, not shown in the top 10 domains in Figure \ref{fig:domain_contribution_network}).
Further breakdown of its performance cost with adPerf reveals two JavaScript documents (\texttt{osd.js} and \texttt{osd\_listener.js}) of size less than $75$ KB belonging to this domain are referenced by over 200 websites in the news corpus.
These two files are part of Google Ads that track the viewability of the ads to assess the value of an impression to the publisher and advertiser.
To calculate what percentage of an ad appears in a viewable space on the screen and for how long that portion of the ad remains visible, these JavaScript snippets are frequently are invoked by the webpage and take up precious CPU cycles.

\textbf{\emph{Breakdown of performance cost by trustworthiness.}}
When an ad is included by the publisher, there is an \textit{explicit} trust between them.
However, when syndication is performed by the ad provider, the ad will be served through a chain of redirections going through different ad domains.
Our measurement result on the Alexa news and general websites shows that the mean depth of the chain is 4, revealing ad syndication is prevalent.
Most of the ad domains on the chain are not directly visible to the publisher (except the one embedded by the publisher) and their intention (e.g., whether they are used for drive-by download or phishing) cannot be verified by the publisher.
There is an \textit{implicit} trust placed by the publisher on the ads but the real trustworthiness of those ad domains is uncertain.
We are interested in the correlation between the performance cost an ad domain brings and its trustworthiness.

To this end, we leveraged two online services, WOT (Web of Trust)~\cite{wot} and VirusTotal~\cite{virustotal}, to model the trustworthiness of an ad domain.
WOT is a community-based reputation system that assigns a score to a domain name based on user complaints and other blacklists.
The score ranges from 0 to 100 and WOT classifies domains based on their scores into $5$ \emph{trust rating} -- excellent, good, unsatisfactory, poor, and very poor \cite{chia2011re}.
VirusTotal is a portal that proxies the request of a security check of a domain/URL to its affiliated blacklist services (71 blacklists).
When a domain is submitted to VirusTotal, it reports the blacklists that flag it as \emph{red}.
We count the ratio of blacklists that did not raise an alarm on the domain (i.e. safe flag) as the VirusTotal score (i.e., 0 means highly malicious and 1 means completely benign).
Both WOT and VirusTotal have been used to determine the trustworthiness of a domain by previous studies~\cite{binsalleeh2014analysis, ikram2019chain, chia2011re}.

\begin{figure} [htbp]
\vspace{-2em}
\def\tabularxcolumn#1{m{#1}}
\begin{tabularx}{.5\textwidth}{cc}

\subfloat[WOT]{\includegraphics[width=0.45\textwidth]{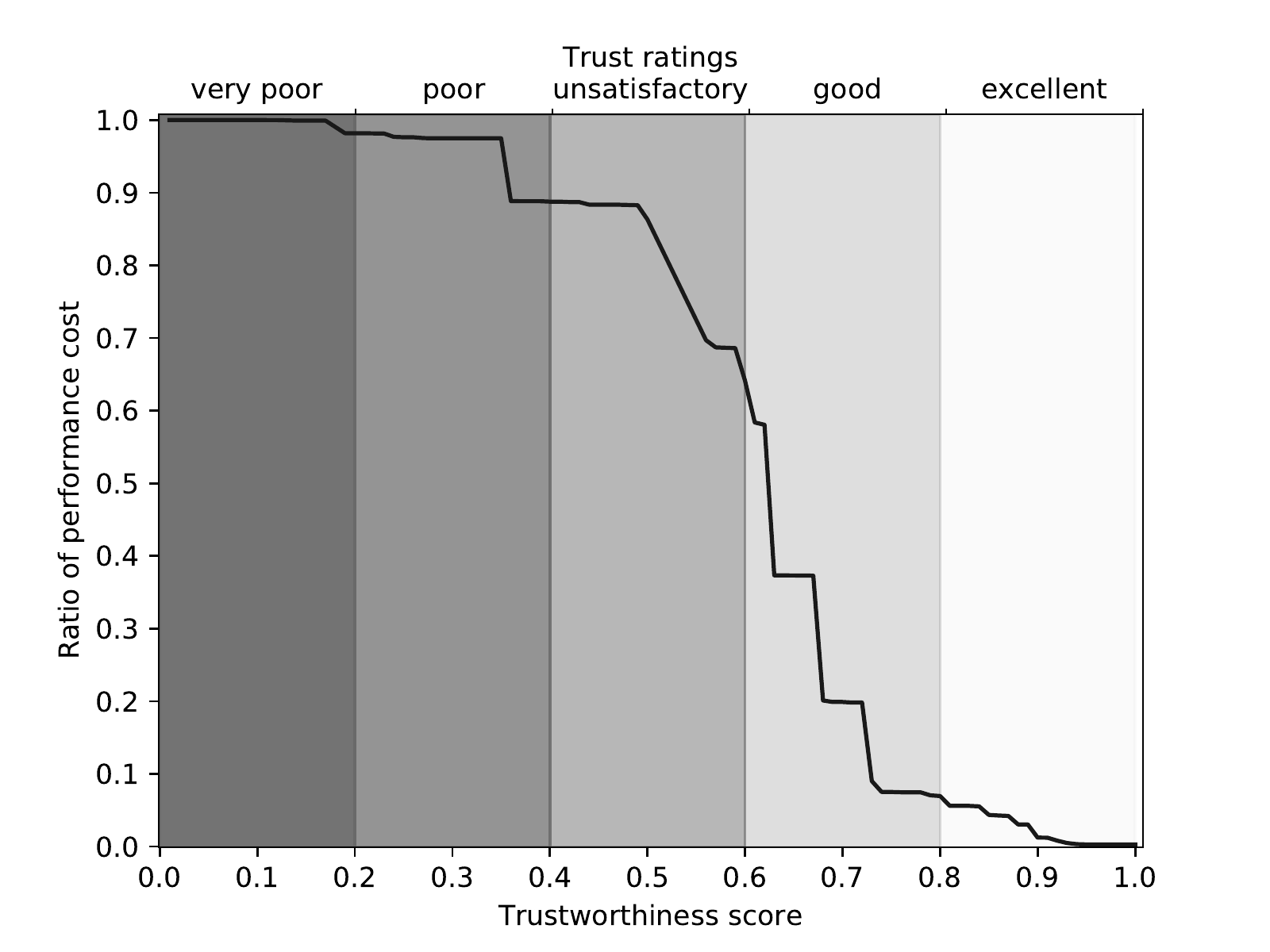}}
\\
\subfloat[VirusTotal]{\includegraphics[width=0.45\textwidth]{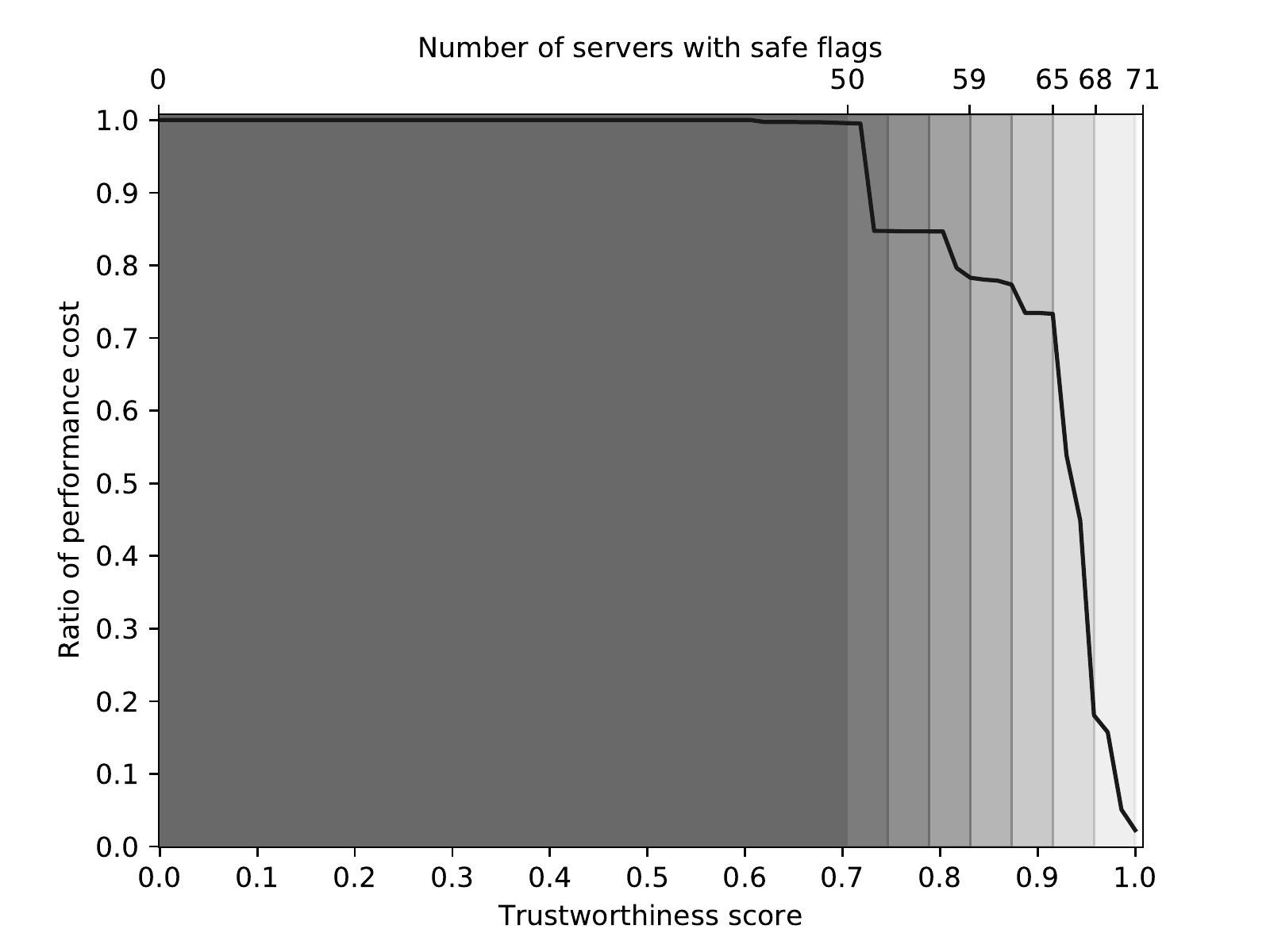}}

\end{tabularx}
\caption{Performance cost of ads delivered by ad domains as a function of its trustworthiness score (CDF). 
Both scores, WOT (top) and VirusTotal (bottom) are normalized to [0,1]. 
Different colors highlight different trustworthiness rating.}\label{fig:trust2}
\vspace{-1em}

\end{figure}

We find that the thresholds for trust ratings vary across different services that report a \emph{trustworthiness score}.
Therefore, to provide a fair analysis, we report the contribution of domains to the ad cost for varying thresholds.
Figure~\ref{fig:trust2} illustrates the cumulative performance cost of ad domains as a function of trustworthiness assessed by WOT and VirusTotal. 
For WOT, we use its default classification ($5$ classes) \cite{chia2011re}. 
For VirusTotal, we observe that almost all of the domains receive at least 50 safe flags, so we only breakdown the region from 50 to 71 servers at the granularity of $3$ servers.

\textbf{Finding 11.} Following the default classification of WOT, about 63\% of ads cost is from ads delivered by trusted ad domains (excellent and good rating).
Nevertheless, domains that are not trusted (unsatisfactory, poor and very poor rating) contribute to a considerable portion of ads (37\%) which is a flag for publishers. 
Accordingly, for VirusTotal, we see that only 5\% of the performance cost of ads is connected to domains that don't receive any red flags.

\textbf{Finding 12.} Domains that are moderately trusted (i.e., neither highly trusted nor untrusted) have the greatest contribution to the performance cost of ads. 
This can be interpreted from Figure~\ref{fig:trust2}.
The amount of drop in the fraction of performance cost (y-axis) within each region indicates the portion of performance cost for that level of trust. 
For example, domains with more than 80\% WOT score (excellent trust rating) contribute to 5\% of ads performance cost while 58\% of ads cost belongs to domains with 60\% to 80\% score (good trust rating). 
Likewise, domains with less than $3$ VirusTotal red flags (first class from the right in the figure) count for 18\% of ads cost but 55\% for domains with $3$ to $6$ red flags (second class from the right). 
Nonetheless, our results do not assert a strong correlation between trustworthiness and the performance cost associated with the third-party ad domains.

\begin{figure} [htbp]
\vspace{-3em}
\def\tabularxcolumn#1{m{#1}}
\begin{tabularx}{.5\textwidth}{cc}

\subfloat[Number of referred websites]{\includegraphics[width=0.44\textwidth]{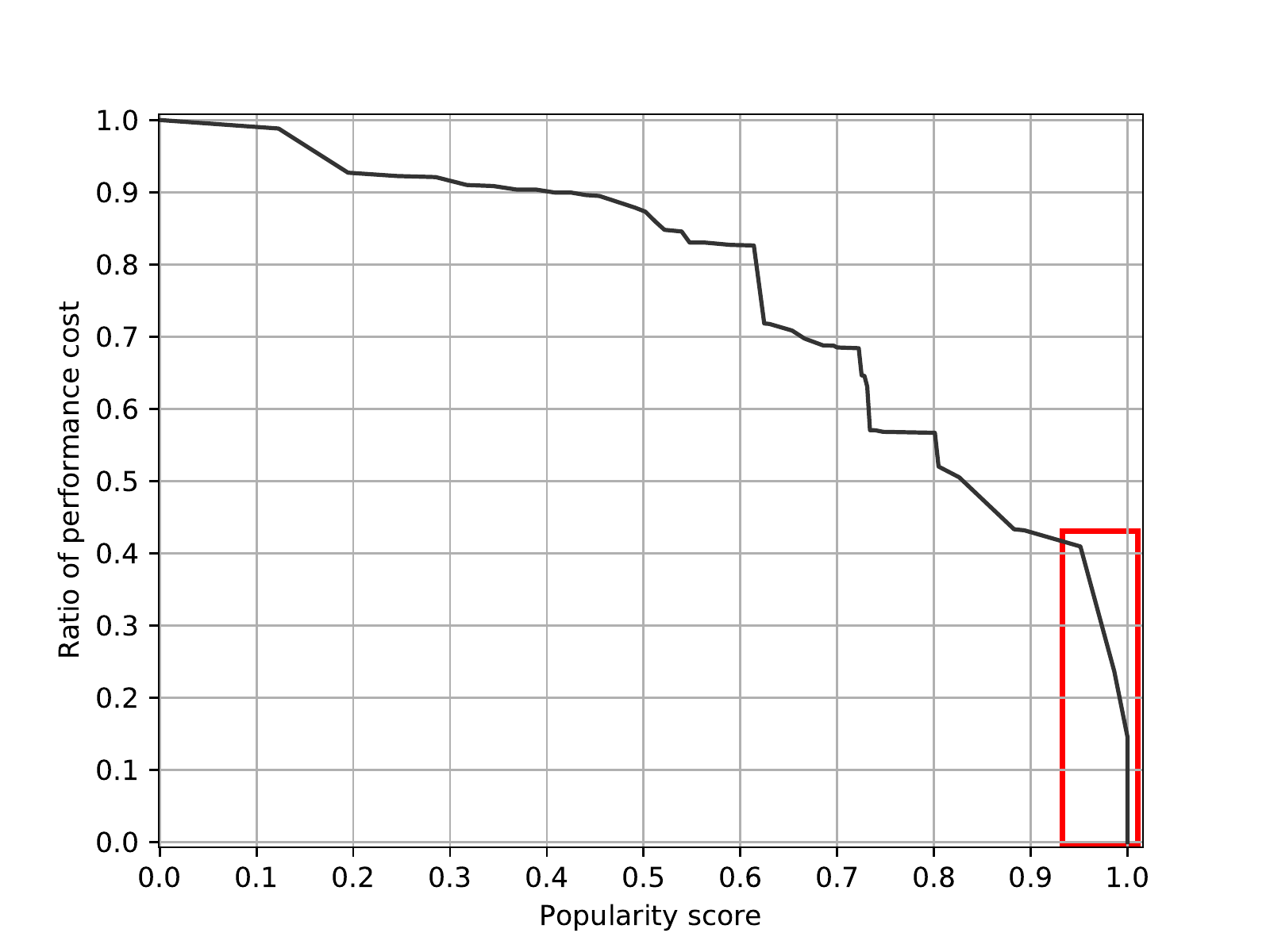}}
\\
\subfloat[Alexa ranking]{\includegraphics[width=0.44\textwidth]{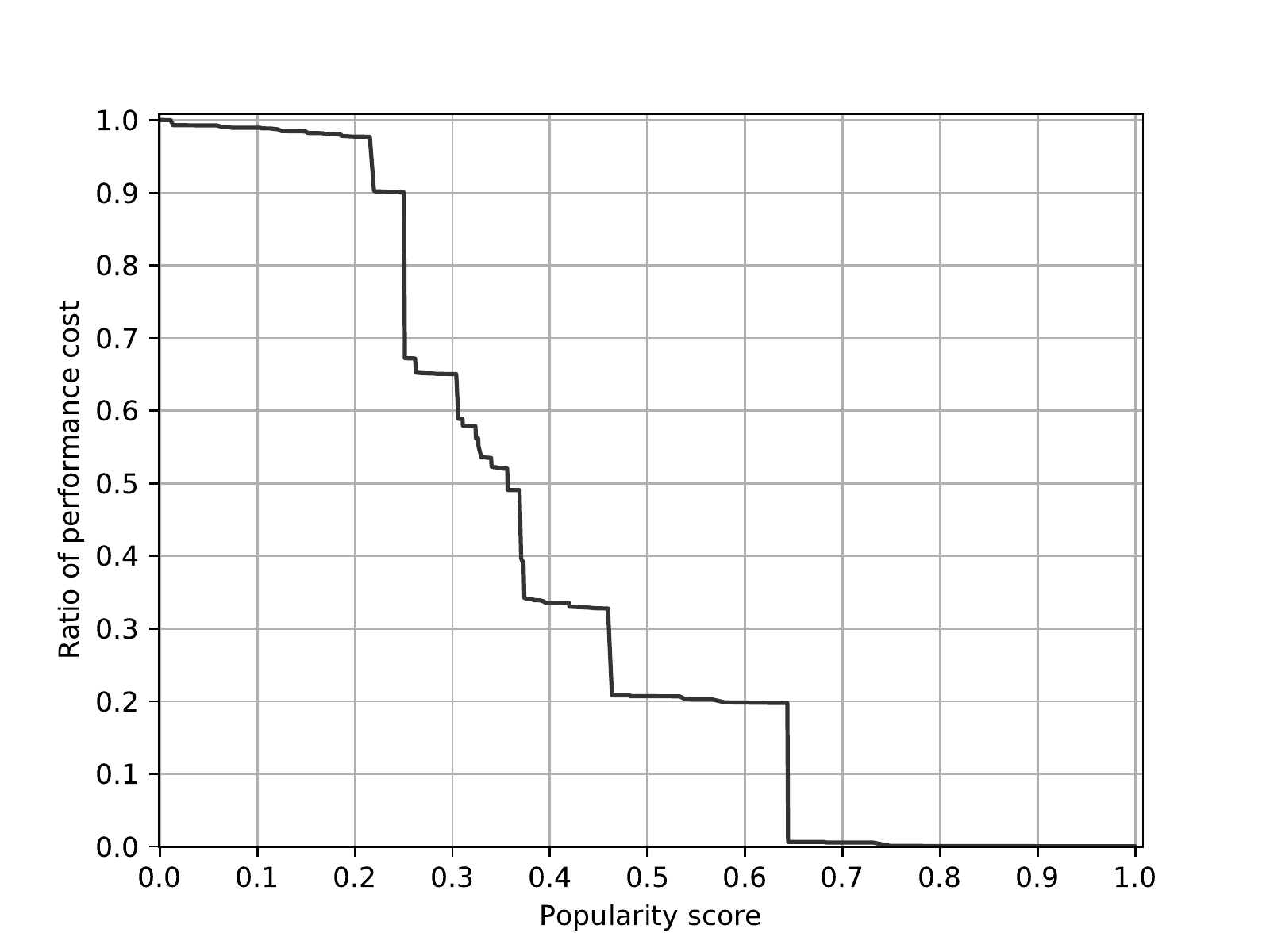}}

\end{tabularx}
\caption{Performance cost of ads from popular domains as a function of popularity score (CDF) on the number of referrers (top) and Alexa ranking (bottom).}\label{fig:popularity}
\vspace{-1em}
\end{figure}

\textbf{\emph{Breakdown of performance cost by popularity.}}
Similar to the trustworthiness gauged by the hygiene of the delivered content, the reputation of an ad domain can be correlated with the performance cost as well.
One might expect the more popular ad domains to contribute a higher fraction of ads cost in the web ecosystem.
To test this hypothesis, we first model the domain reputation by its popularity, which is determined by the Alexa ranking~\cite{alexa} and the number of websites in our corpus referring to it.
However, there is no agreed-upon cutoff to split ad domains into popular versus unpopular.
For this reason, we follow a similar method to the trustworthiness study and draw the contribution of popular ad domains to the performance cost at varying cutoff levels.
Figure~\ref{fig:popularity} illustrates the cumulative contribution of popular domains to the performance cost of ads for two metrics.

\textbf{Finding 13.} Even though earlier in this section we observed no correlation between the popularity of the ad domains (i.e. number of referred websites) and the performance cost for multiple domains, at the macro-level, more popular ad domains contribute more to the performance cost as we can see from Figure~\ref{fig:popularity}(a).
As highligted in this figure, the fraction of performance cost drops about 40\% within 5\% range of the most popular ad domains.  
However, for the Alexa ranking, we observe multiple sharp drops throughout the score range, meaning there exist multiple ad-domains that have a significant contribution to the performance that is neither very popular nor very unpopular.

\vspace{-0.07cm}
\section{Conclusions and Takeaways}
\label{sec:conc}
Our evaluations on the performance cost of ads lead to multiple new and interesting observations.
The key finding of this research is ads have significant cost, more than 15\% of page loading computation workload.
In addition to this, we discover Scripting contributes to more than 88\% of this cost.
We also find out that the breakdown of activities in displaying ads does not completely align with the main contents of the pages.
For example, HTML parsing comprises of more ad-contents rather than the main content.
Web ads have also increased the time spent on the network by almost the same ratio.
About 80\% of the time overhead of ads is related to images and script resources.
Also, our results show that resources related to ad content are relatively dissimilar to the typical resources requested by pages.
For example, XML files are requested more by the ad contents.
Our evaluations also show that a considerable fraction of the performance cost of ads is from untrusted domains.

We design and implement adPerf to provide insight and guidance to both publishers and third-party ad providers to improve the performance of ads by identifying the stage and/or resource which are the main bottlenecks.
For example, if adPerf identifies Scripting to be the computation and network bottleneck of ads on a website, one can follow targeted optimizations to loading third-party JavaScript provided by Google Page Insights \cite{recom} such as lazy-loading scripts and libraries (e.g. serving an ad in the footer only when a user scrolls down the page), splitting JavaScript bundles (e.g. dynamic import() statement), self-hosting scripts with Service workers particularly for ad domains with consistent APIs, using resource hints like preconnect and DNS-prefetch, sandboxing script with iframes, and using asynchronous ad tag manager in the code to name a few.

In this study, we did not account for ad resources that might be directly embedded in native HTML and ad resources that cannot be detected by filter lists (i.e. websites that use circumvention to evade filter lists).
In future work, we plan to also include such sources of ad content.
This addition would only increase the performance costs of different ad breakdowns reported throughout this paper, which we believe is already significant to warrant deeper attention.
We anticipate this study primarily aimed at designing a methodology and open-source infrastructure for fine-grained analysis of ads will be a useful tool for web researchers to prioritize their optimization efforts and publishers to analyze the impact of their ad-delivery chain.

\bibliographystyle{ACM-Reference-Format}
\balance
\bibliography{main}


\begin{thebibliography}{40}


\ifx \showCODEN    \undefined \def \showCODEN     #1{\unskip}     \fi
\ifx \showDOI      \undefined \def \showDOI       #1{#1}\fi
\ifx \showISBNx    \undefined \def \showISBNx     #1{\unskip}     \fi
\ifx \showISBNxiii \undefined \def \showISBNxiii  #1{\unskip}     \fi
\ifx \showISSN     \undefined \def \showISSN      #1{\unskip}     \fi
\ifx \showLCCN     \undefined \def \showLCCN      #1{\unskip}     \fi
\ifx \shownote     \undefined \def \shownote      #1{#1}          \fi
\ifx \showarticletitle \undefined \def \showarticletitle #1{#1}   \fi
\ifx \showURL      \undefined \def \showURL       {\relax}        \fi
\providecommand\bibfield[2]{#2}
\providecommand\bibinfo[2]{#2}
\providecommand\natexlab[1]{#1}
\providecommand\showeprint[2][]{arXiv:#2}

\bibitem[\protect\citeauthoryear{??}{iab}{2020}]%
        {iab}
 \bibinfo{year}{2020}\natexlab{}.
\newblock \bibinfo{title}{{AD BLOCK DETECTION SCRIPT}}.
\newblock
  \bibinfo{howpublished}{\url{https://iabtechlab.com/software/ad-block-detection-script/}}.
\newblock


\bibitem[\protect\citeauthoryear{??}{adb}{2020a}]%
        {adblockplus}
 \bibinfo{year}{2020}\natexlab{a}.
\newblock \bibinfo{title}{{Adblock Plus}}.
\newblock \bibinfo{howpublished}{\url{https://adblockplus.org}}.
\newblock


\bibitem[\protect\citeauthoryear{??}{adb}{2020b}]%
        {adblockparser}
 \bibinfo{year}{2020}\natexlab{b}.
\newblock \bibinfo{title}{{adblockparser}}.
\newblock
  \bibinfo{howpublished}{\url{https://github.com/scrapinghub/adblockparser}}.
\newblock


\bibitem[\protect\citeauthoryear{??}{ale}{2020a}]%
        {alexa}
 \bibinfo{year}{2020}\natexlab{a}.
\newblock \bibinfo{title}{{Alexa Top News Sites}}.
\newblock
  \bibinfo{howpublished}{\url{https://www.alexa.com/topsites/category/News}}.
\newblock


\bibitem[\protect\citeauthoryear{??}{ale}{2020b}]%
        {alexaus}
 \bibinfo{year}{2020}\natexlab{b}.
\newblock \bibinfo{title}{{Alexa Top Sites}}.
\newblock
  \bibinfo{howpublished}{\url{https://www.alexa.com/topsites/countries/US}}.
\newblock


\bibitem[\protect\citeauthoryear{??}{pro}{2020}]%
        {protocol}
 \bibinfo{year}{2020}\natexlab{}.
\newblock \bibinfo{title}{{Chrome DevTools Protocol}}.
\newblock
  \bibinfo{howpublished}{\url{https://chromedevtools.github.io/devtools-protocol}}.
\newblock


\bibitem[\protect\citeauthoryear{??}{eas}{2020}]%
        {easylist}
 \bibinfo{year}{2020}\natexlab{}.
\newblock \bibinfo{title}{{EasyList}}.
\newblock \bibinfo{howpublished}{\url{https://easylist.to}}.
\newblock


\bibitem[\protect\citeauthoryear{??}{Gec}{2020}]%
        {Gecko-profiler}
 \bibinfo{year}{2020}\natexlab{}.
\newblock \bibinfo{title}{Gecko profiler}.
\newblock
\newblock
\newblock
\shownote{\url{https://developer.mozilla.org/en-US/docs/Mozilla/Performance/Profiling_with_the_Built-in_Profiler}.}


\bibitem[\protect\citeauthoryear{??}{rec}{2020}]%
        {recom}
 \bibinfo{year}{2020}\natexlab{}.
\newblock \bibinfo{title}{{Loading Third-Party JavaScript}}.
\newblock
  \bibinfo{howpublished}{\url{https://developers.google.com/web/fundamentals/performance/optimizing-content-efficiency/loading-third-party-javascript/?utm_source=lighthouse&utm_medium=unknown}}.
\newblock


\bibitem[\protect\citeauthoryear{??}{pop}{2020}]%
        {popet}
 \bibinfo{year}{2020}\natexlab{}.
\newblock \bibinfo{title}{{Popeteer}}.
\newblock
  \bibinfo{howpublished}{\url{https://developers.google.com/web/tools/puppeteer/get-started}}.
\newblock


\bibitem[\protect\citeauthoryear{??}{Chr}{2020}]%
        {Chromeprofiler}
 \bibinfo{year}{2020}\natexlab{}.
\newblock \bibinfo{title}{The Trace Event Profiling Tool}.
\newblock
  \bibinfo{howpublished}{https://www.chromium.org/developers/how-tos/trace-event-profiling-tool}.
\newblock


\bibitem[\protect\citeauthoryear{??}{vir}{2020}]%
        {virustotal}
 \bibinfo{year}{2020}\natexlab{}.
\newblock \bibinfo{title}{{VirusTotal}}.
\newblock \bibinfo{howpublished}{\url{https://www.virustotal.com}}.
\newblock


\bibitem[\protect\citeauthoryear{??}{wot}{2020}]%
        {wot}
 \bibinfo{year}{2020}\natexlab{}.
\newblock \bibinfo{title}{{Website Safety, Security Check Web Of Trust}}.
\newblock \bibinfo{howpublished}{\url{https://www.mywot.com/}}.
\newblock


\bibitem[\protect\citeauthoryear{??}{zbr}{2020}]%
        {zbrowse}
 \bibinfo{year}{2020}\natexlab{}.
\newblock \bibinfo{title}{Zbrowse}.
\newblock \bibinfo{howpublished}{\url{https://github.com/zmap/zbrowse}}.
\newblock


\bibitem[\protect\citeauthoryear{An}{An}{2018}]%
        {An:2018aa}
\bibfield{author}{\bibinfo{person}{Daniel An}.}
  \bibinfo{year}{2018}\natexlab{}.
\newblock \bibinfo{title}{Mobile page speed}.
\newblock
  \bibinfo{howpublished}{\url{https://www.thinkwithgoogle.com/marketing-resources/data-measurement/mobile-page-speed-new-industry-benchmarks/}}.
\newblock


\bibitem[\protect\citeauthoryear{Anthes}{Anthes}{[n. d.]}]%
        {brokers}
\bibfield{author}{\bibinfo{person}{Gary Anthes}.} \bibinfo{year}{[n.
  d.]}\natexlab{}.
\newblock \showarticletitle{Data brokers are watching you.}
\newblock \bibinfo{journal}{\emph{Commun. ACM}} \bibinfo{volume}{58},
  \bibinfo{number}{1} (\bibinfo{year}{[n. d.]}), \bibinfo{pages}{28--30}.
\newblock


\bibitem[\protect\citeauthoryear{Bashir, Arshad, Robertson, and Wilson}{Bashir
  et~al\mbox{.}}{2016}]%
        {bashir2016tracing}
\bibfield{author}{\bibinfo{person}{Muhammad~Ahmad Bashir},
  \bibinfo{person}{Sajjad Arshad}, \bibinfo{person}{William Robertson}, {and}
  \bibinfo{person}{Christo Wilson}.} \bibinfo{year}{2016}\natexlab{}.
\newblock \showarticletitle{Tracing information flows between ad exchanges
  using retargeted ads}. In \bibinfo{booktitle}{\emph{25th {USENIX} Security
  Symposium}}. \bibinfo{pages}{481--496}.
\newblock


\bibitem[\protect\citeauthoryear{Binsalleeh}{Binsalleeh}{2014}]%
        {binsalleeh2014analysis}
\bibfield{author}{\bibinfo{person}{Hamad Binsalleeh}.}
  \bibinfo{year}{2014}\natexlab{}.
\newblock \emph{\bibinfo{title}{Analysis of Malware and Domain Name System
  Traffic}}.
\newblock \bibinfo{thesistype}{Ph.D. Dissertation}. \bibinfo{school}{Concordia
  University}.
\newblock


\bibitem[\protect\citeauthoryear{Chia and Knapskog}{Chia and Knapskog}{2011}]%
        {chia2011re}
\bibfield{author}{\bibinfo{person}{Pern~Hui Chia} {and}
  \bibinfo{person}{Svein~Johan Knapskog}.} \bibinfo{year}{2011}\natexlab{}.
\newblock \showarticletitle{Re-evaluating the wisdom of crowds in assessing web
  security}. In \bibinfo{booktitle}{\emph{International Conference on Financial
  Cryptography and Data Security}}. Springer, \bibinfo{pages}{299--314}.
\newblock


\bibitem[\protect\citeauthoryear{Curtsinger and Berger}{Curtsinger and
  Berger}{2015}]%
        {cozpaper}
\bibfield{author}{\bibinfo{person}{Charlie Curtsinger} {and}
  \bibinfo{person}{Emery~D Berger}.} \bibinfo{year}{2015}\natexlab{}.
\newblock \showarticletitle{COZ: Finding Code that Counts with Causal
  Profiling}. In \bibinfo{booktitle}{\emph{Proceedings of the 25th Symposium on
  Operating Systems Principles}}. ACM, \bibinfo{pages}{184--197}.
\newblock


\bibitem[\protect\citeauthoryear{De~Corniere and De~Nijs}{De~Corniere and
  De~Nijs}{2016}]%
        {onlineads}
\bibfield{author}{\bibinfo{person}{Alexandre De~Corniere} {and}
  \bibinfo{person}{Romain De~Nijs}.} \bibinfo{year}{2016}\natexlab{}.
\newblock \showarticletitle{Online advertising and privacy}.
\newblock \bibinfo{journal}{\emph{The RAND Journal of Economics}}
  \bibinfo{volume}{47}, \bibinfo{number}{1} (\bibinfo{year}{2016}),
  \bibinfo{pages}{48--72}.
\newblock


\bibitem[\protect\citeauthoryear{Englehardt and Narayanan}{Englehardt and
  Narayanan}{2016}]%
        {oneMilion}
\bibfield{author}{\bibinfo{person}{Steven Englehardt} {and}
  \bibinfo{person}{Arvind Narayanan}.} \bibinfo{year}{2016}\natexlab{}.
\newblock \showarticletitle{Online tracking: A 1-million-site measurement and
  analysis}. In \bibinfo{booktitle}{\emph{Proceedings of the 2016 ACM SIGSAC
  conference on computer and communications security}}. ACM,
  \bibinfo{pages}{1388--1401}.
\newblock


\bibitem[\protect\citeauthoryear{Evans}{Evans}{2009}]%
        {onlineads1}
\bibfield{author}{\bibinfo{person}{David~S Evans}.}
  \bibinfo{year}{2009}\natexlab{}.
\newblock \showarticletitle{The online advertising industry: Economics,
  evolution, and privacy}.
\newblock \bibinfo{journal}{\emph{Journal of economic perspectives}}
  \bibinfo{volume}{23}, \bibinfo{number}{3} (\bibinfo{year}{2009}),
  \bibinfo{pages}{37--60}.
\newblock


\bibitem[\protect\citeauthoryear{Garimella, Kostakis, and
  Mathioudakis}{Garimella et~al\mbox{.}}{2017}]%
        {adblockPerformance}
\bibfield{author}{\bibinfo{person}{Kiran Garimella}, \bibinfo{person}{Orestis
  Kostakis}, {and} \bibinfo{person}{Michael Mathioudakis}.}
  \bibinfo{year}{2017}\natexlab{}.
\newblock \showarticletitle{Ad-blocking: A study on performance, privacy and
  counter-measures}. In \bibinfo{booktitle}{\emph{Proceedings of the 2017 ACM
  on Web Science Conference}}. ACM, \bibinfo{pages}{259--262}.
\newblock


\bibitem[\protect\citeauthoryear{Golestani, Mahlke, and Narayanasamy}{Golestani
  et~al\mbox{.}}{2019}]%
        {golestani}
\bibfield{author}{\bibinfo{person}{Hossein Golestani}, \bibinfo{person}{Scott
  Mahlke}, {and} \bibinfo{person}{Satish Narayanasamy}.}
  \bibinfo{year}{2019}\natexlab{}.
\newblock \showarticletitle{Characterization of Unnecessary Computations in Web
  Applications}. In \bibinfo{booktitle}{\emph{2019 IEEE International Symposium
  on Performance Analysis of Systems and Software (ISPASS)}}. IEEE,
  \bibinfo{pages}{11--21}.
\newblock


\bibitem[\protect\citeauthoryear{Ikram and Kaafar}{Ikram and Kaafar}{2017}]%
        {ikram2017first}
\bibfield{author}{\bibinfo{person}{Muhammad Ikram} {and}
  \bibinfo{person}{Mohamed~Ali Kaafar}.} \bibinfo{year}{2017}\natexlab{}.
\newblock \showarticletitle{A first look at mobile ad-blocking apps}. In
  \bibinfo{booktitle}{\emph{2017 IEEE 16th International Symposium on Network
  Computing and Applications (NCA)}}. IEEE, \bibinfo{pages}{1--8}.
\newblock


\bibitem[\protect\citeauthoryear{Ikram, Masood, Tyson, Kaafar, Loizon, and
  Ensafi}{Ikram et~al\mbox{.}}{2019}]%
        {ikram2019chain}
\bibfield{author}{\bibinfo{person}{Muhammad Ikram}, \bibinfo{person}{Rahat
  Masood}, \bibinfo{person}{Gareth Tyson}, \bibinfo{person}{Mohamed~Ali
  Kaafar}, \bibinfo{person}{Noha Loizon}, {and} \bibinfo{person}{Roya Ensafi}.}
  \bibinfo{year}{2019}\natexlab{}.
\newblock \showarticletitle{The chain of implicit trust: {An} analysis of the
  web third-party resources loading}. In \bibinfo{booktitle}{\emph{The World
  Wide Web Conference}}. ACM, \bibinfo{pages}{2851--2857}.
\newblock


\bibitem[\protect\citeauthoryear{Iqbal, Shafiq, and Qian}{Iqbal
  et~al\mbox{.}}{2017}]%
        {iqbal2017ad}
\bibfield{author}{\bibinfo{person}{Umar Iqbal}, \bibinfo{person}{Zubair
  Shafiq}, {and} \bibinfo{person}{Zhiyun Qian}.}
  \bibinfo{year}{2017}\natexlab{}.
\newblock \showarticletitle{The ad wars: retrospective measurement and analysis
  of anti-adblock filter lists}. In \bibinfo{booktitle}{\emph{Proceedings of
  the 2017 Internet Measurement Conference}}. ACM, \bibinfo{pages}{171--183}.
\newblock


\bibitem[\protect\citeauthoryear{Iqbal, Shafiq, Snyder, Zhu, Qian, and
  Livshits}{Iqbal et~al\mbox{.}}{2018}]%
        {iqbal2018adgraph}
\bibfield{author}{\bibinfo{person}{Umar Iqbal}, \bibinfo{person}{Zubair
  Shafiq}, \bibinfo{person}{Peter Snyder}, \bibinfo{person}{Shitong Zhu},
  \bibinfo{person}{Zhiyun Qian}, {and} \bibinfo{person}{Benjamin Livshits}.}
  \bibinfo{year}{2018}\natexlab{}.
\newblock \showarticletitle{Adgraph: A machine learning approach to automatic
  and effective adblocking}.
\newblock \bibinfo{journal}{\emph{arXiv preprint arXiv:1805.09155}}
  (\bibinfo{year}{2018}).
\newblock


\bibitem[\protect\citeauthoryear{Lerner, Simpson, Kohno, and Roesner}{Lerner
  et~al\mbox{.}}{2016}]%
        {trackingStudy}
\bibfield{author}{\bibinfo{person}{Adam Lerner}, \bibinfo{person}{Anna~Kornfeld
  Simpson}, \bibinfo{person}{Tadayoshi Kohno}, {and} \bibinfo{person}{Franziska
  Roesner}.} \bibinfo{year}{2016}\natexlab{}.
\newblock \showarticletitle{Internet jones and the raiders of the lost
  trackers: An archaeological study of web tracking from 1996 to 2016}. In
  \bibinfo{booktitle}{\emph{25th $\{$USENIX$\}$ Security Symposium
  ($\{$USENIX$\}$ Security 16)}}.
\newblock


\bibitem[\protect\citeauthoryear{Li, Zhang, Xie, Yu, and Wang}{Li
  et~al\mbox{.}}{2012}]%
        {li2012knowing}
\bibfield{author}{\bibinfo{person}{Zhou Li}, \bibinfo{person}{Kehuan Zhang},
  \bibinfo{person}{Yinglian Xie}, \bibinfo{person}{Fang Yu}, {and}
  \bibinfo{person}{XiaoFeng Wang}.} \bibinfo{year}{2012}\natexlab{}.
\newblock \showarticletitle{Knowing your enemy: understanding and detecting
  malicious web advertising}. In \bibinfo{booktitle}{\emph{Proceedings of the
  2012 ACM conference on Computer and communications security}}. ACM,
  \bibinfo{pages}{674--686}.
\newblock


\bibitem[\protect\citeauthoryear{Meyerovich and Bodik}{Meyerovich and
  Bodik}{2010}]%
        {fastlayout}
\bibfield{author}{\bibinfo{person}{Leo~A Meyerovich} {and}
  \bibinfo{person}{Rastislav Bodik}.} \bibinfo{year}{2010}\natexlab{}.
\newblock \showarticletitle{Fast and parallel webpage layout}. In
  \bibinfo{booktitle}{\emph{Proceedings of the 19th international conference on
  World wide web}}. ACM, \bibinfo{pages}{711--720}.
\newblock


\bibitem[\protect\citeauthoryear{Mughees, Qian, and Shafiq}{Mughees
  et~al\mbox{.}}{2017}]%
        {mughees2017detecting}
\bibfield{author}{\bibinfo{person}{Muhammad~Haris Mughees},
  \bibinfo{person}{Zhiyun Qian}, {and} \bibinfo{person}{Zubair Shafiq}.}
  \bibinfo{year}{2017}\natexlab{}.
\newblock \showarticletitle{Detecting anti ad-blockers in the wild}.
\newblock \bibinfo{journal}{\emph{Proceedings on Privacy Enhancing
  Technologies}} \bibinfo{volume}{2017}, \bibinfo{number}{3}
  (\bibinfo{year}{2017}), \bibinfo{pages}{130--146}.
\newblock


\bibitem[\protect\citeauthoryear{Nejati and Balasubramanian}{Nejati and
  Balasubramanian}{2016}]%
        {wprofm}
\bibfield{author}{\bibinfo{person}{Javad Nejati} {and} \bibinfo{person}{Aruna
  Balasubramanian}.} \bibinfo{year}{2016}\natexlab{}.
\newblock \showarticletitle{An {I}n-depth study of {M}obile {B}rowser
  {P}erformance}. In \bibinfo{booktitle}{\emph{Proc. of the 25th Intl. Conf. on
  WWW}}. Intl. WWW Conf. Steering Committee, \bibinfo{pages}{1305--1315}.
\newblock


\bibitem[\protect\citeauthoryear{Pourghassemi, Sani, and
  Chandramowlishwaran}{Pourghassemi et~al\mbox{.}}{2019}]%
        {Pourghassemi:2019aa}
\bibfield{author}{\bibinfo{person}{Behnam Pourghassemi},
  \bibinfo{person}{Ardalan~Amiri Sani}, {and} \bibinfo{person}{Aparna
  Chandramowlishwaran}.} \bibinfo{year}{2019}\natexlab{}.
\newblock \showarticletitle{What-If Analysis of Page Load Time in Web Browsers
  Using Causal Profiling}.
\newblock \bibinfo{journal}{\emph{Proceedings of the ACM on Measurement and
  Analysis of Computing Systems (POMACS)}} (\bibinfo{date}{June}
  \bibinfo{year}{2019}).
\newblock
\newblock
\shownote{(SIGMETRICS).}


\bibitem[\protect\citeauthoryear{Pujol, Hohlfeld, and Feldmann}{Pujol
  et~al\mbox{.}}{2015}]%
        {annoyedUsers}
\bibfield{author}{\bibinfo{person}{Enric Pujol}, \bibinfo{person}{Oliver
  Hohlfeld}, {and} \bibinfo{person}{Anja Feldmann}.}
  \bibinfo{year}{2015}\natexlab{}.
\newblock \showarticletitle{Annoyed users: Ads and ad-block usage in the wild}.
  In \bibinfo{booktitle}{\emph{Proceedings of the 2015 Internet Measurement
  Conference}}. ACM, \bibinfo{pages}{93--106}.
\newblock


\bibitem[\protect\citeauthoryear{Wang, Balasubramanian, Krishnamurthy, and
  Wetherall}{Wang et~al\mbox{.}}{2013}]%
        {wprof}
\bibfield{author}{\bibinfo{person}{Xiao~Sophia Wang}, \bibinfo{person}{Aruna
  Balasubramanian}, \bibinfo{person}{Arvind Krishnamurthy}, {and}
  \bibinfo{person}{David Wetherall}.} \bibinfo{year}{2013}\natexlab{}.
\newblock \showarticletitle{Demystifying Page Load Performance with WProf.}. In
  \bibinfo{booktitle}{\emph{NSDI}}. \bibinfo{pages}{473--485}.
\newblock


\bibitem[\protect\citeauthoryear{Wang, Lin, Zhong, and Chishtie}{Wang
  et~al\mbox{.}}{2011}]%
        {browsers-slow-smartphones}
\bibfield{author}{\bibinfo{person}{Zhen Wang}, \bibinfo{person}{Felix~Xiaozhu
  Lin}, \bibinfo{person}{Lin Zhong}, {and} \bibinfo{person}{Mansoor Chishtie}.}
  \bibinfo{year}{2011}\natexlab{}.
\newblock \showarticletitle{Why {A}re {W}eb {B}rowsers {S}low on
  {S}martphones?}. In \bibinfo{booktitle}{\emph{Proceedings of the 12th
  Workshop on Mobile Computing Systems and Applications}}
  \emph{(\bibinfo{series}{HotMobile '11})}. \bibinfo{publisher}{ACM},
  \bibinfo{address}{New York, NY, USA}.
\newblock


\bibitem[\protect\citeauthoryear{Yu, Macbeth, Modi, and Pujol}{Yu
  et~al\mbox{.}}{2016}]%
        {tackingthetrackers}
\bibfield{author}{\bibinfo{person}{Zhonghao Yu}, \bibinfo{person}{Sam Macbeth},
  \bibinfo{person}{Konark Modi}, {and} \bibinfo{person}{Josep~M Pujol}.}
  \bibinfo{year}{2016}\natexlab{}.
\newblock \showarticletitle{Tracking the trackers}. In
  \bibinfo{booktitle}{\emph{Proceedings of the 25th International Conference on
  World Wide Web}}. International World Wide Web Conferences Steering
  Committee, \bibinfo{pages}{121--132}.
\newblock


\bibitem[\protect\citeauthoryear{Zhu, Iqbal, Wang, Qian, Shafiq, and Chen}{Zhu
  et~al\mbox{.}}{2019}]%
        {zhu2019shadowblock}
\bibfield{author}{\bibinfo{person}{Shitong Zhu}, \bibinfo{person}{Umar Iqbal},
  \bibinfo{person}{Zhongjie Wang}, \bibinfo{person}{Zhiyun Qian},
  \bibinfo{person}{Zubair Shafiq}, {and} \bibinfo{person}{Weiteng Chen}.}
  \bibinfo{year}{2019}\natexlab{}.
\newblock \showarticletitle{ShadowBlock: A Lightweight and Stealthy Adblocking
  Browser}. In \bibinfo{booktitle}{\emph{The World Wide Web Conference}}. ACM,
  \bibinfo{pages}{2483--2493}.
\newblock


\end{thebibliography}

\end{document}